\documentclass[aps,prd,superscriptaddress,longbibliography,nofootinbib,preprintnumbers,reprint]{revtex4-2}
\usepackage[utf8]{inputenc} 
\usepackage[T1]{fontenc} 
\usepackage{lmodern} 
\usepackage{microtype} 
\usepackage[english]{babel}
\usepackage{graphicx}
\usepackage{amsmath}
\usepackage{amssymb}
\usepackage{mathrsfs}
\usepackage{slashed}
\usepackage[usenames,dvipsnames]{xcolor}
\usepackage[colorlinks=true, linkcolor=RoyalBlue, citecolor=RoyalBlue, urlcolor=RoyalBlue]{hyperref}

\def\f{\frac}
\def\mc{\mathcal}

\def\ie{\emph{i.e.}}

\def\v[#1]{\textbf{#1}}
\def\w[#1]{\widehat{#1}}
\def\vs[#1,#2]{\boldsymbol{{#1}_{#2}}}
\def\mes[#1]{d^{3}{#1}}
\def\del{\partial}
\def\<{\langle}
\def\>{\rangle}
\def\vecs[#1,#2]{\boldsymbol{{#1}_{#2}}}

\newcommand{\be}{\begin{equation}}
\newcommand{\ee}{\end{equation}}
\newcommand{\bes}{\begin{subequations}}
\newcommand{\ees}{\end{subequations}}

\newcommand{\half}{\frac{1}{2}}

\def\a{\alpha}
\def\b{\beta}
\def\d{\delta}
\def\D{\Delta}
\def\e{\epsilon}

\def\k{\kappa}

\def\m{\mu}
\def\n{\nu}
\def\N{\nabla}

\def\O{\Omega}

\def\s{\sigma}
\def\t{\tau}
\def\th{\theta}

\begin{document}

\title{Heavy-ion collisions, Gubser flow, and Carroll hydrodynamics}

\author{Arjun Bagchi} 
\email{abagchi@iitk.ac.in}
\affiliation{Indian Institute of Technology Kanpur, Kanpur 208016, India} 
\author{Kedar S.~Kolekar}
\email{kedarsk@mail.tsinghua.edu.cn}
\affiliation{Indian Institute of Technology Kanpur, Kanpur 208016, India}
\affiliation{Yau Mathematical Sciences Center, Tsinghua University, Beijing 100084, China}
\author{Taniya Mandal}
\email{taniyam@iitk.ac.in}
\affiliation{Indian Institute of Technology Kanpur, Kanpur 208016, India}
\author{Ashish Shukla}
\email{ashish.shukla@polytechnique.edu}
\affiliation{CPHT, CNRS, \'Ecole polytechnique, Institut Polytechnique de Paris, 91120 Palaiseau, France}
\affiliation{School of Physics, Korea Institute for Advanced Study, Seoul 02455, South Korea}

\preprint{CPHT-RR046.072023}

\begin{abstract}
Gubser flow provides an analytic model for describing the spacetime dynamics of the quark-gluon plasma produced in heavy-ion collisions. Along with boost and rotation invariance along the beam axis, the model assumes invariance under a combination of translations and special conformal transformations in the transverse plane, leading to a flow profile which evolves not just along the beam axis, but also radially. 
We argue that Gubser flow and its associated symmetry assumptions arise naturally as a consequence of Carrollian symmetries for a conformal Carroll fluid, thereby providing a dual geometric picture for the flow. Given the inherent ultrarelativistic nature of the flow, this duality with Carroll hydrodynamics - which arises in the $c\to 0$ limit of relativistic hydrodynamics, is natural. We provide a precise map between Gubser flow and the conformal Carroll fluid, appropriate to capture the duality between the two not just at the ideal level, but also with the inclusion of hydrodynamic derivative corrections. 
\end{abstract}

\maketitle

\section{Introduction}
\label{sec:intro}
A pervading theme in modern high-energy physics is the investigation of the exotic state of matter that filled the universe ephemerally after the hot Big Bang. This primordial state is known as the Quark-Gluon Plasma (QGP), owing to the fact that the quarks and gluons that constitute ordinary hadronic matter around us existed at these early moments after the Big Bang in a deconfined plasma state. Confinement of the colour charge is an important feature of Quantum Chromodynamics (QCD), the theory describing the strong interactions between quarks and gluons, both of which carry a colour charge. It is at the extreme energy scales present at the birth of the universe that the conditions were right for QCD to display asymptotic freedom, leading to the deconfinement of quarks and gluons and appearance of the QGP. 

Needless to say, it is of immense importance to examine the physical properties of the QGP, to better understand the state of the very early universe and its subsequent evolution, as well as to test QCD under limiting conditions. Experiments such as the Relativistic Heavy Ion Collider (RHIC) at the Brookhaven National Laboratory and A Large Ion Collider Experiment (ALICE) at the Conseil Europ\'een pour la Recherche Nucl\'eaire (CERN) recreate the extreme conditions that were present in the very early universe by colliding highly energetic heavy-ions at ultrarelativistic speeds. When the collision occurs, the local energy density and temperatures produced are high enough to ``melt'' the constituent protons and neutrons of the heavy nuclei, leading to the formation of QGP. A profound outcome of these experiments has been the surprising realization that QGP behaves as a nearly perfect fluid, with the ratio of shear viscosity to entropy density, $\eta/s$, being very small \cite{Gale:2013da}.

A prominent hydrodynamic model that attempts to describe the collective flow of the QGP produced after the heavy-ion collision was proposed by Bjorken \cite{Bjorken:1982qr}. The model, known as \emph{Bjorken flow}, predicts the evolution of the energy density, temperature etc.~for the QGP as a function of the proper time $\tau$, by making certain simplifying assumptions. These include the premise that the flow profile is boost (more precisely, rapidity) independent along the beam axis, as well as translation and rotation invariant in the plane transverse to the beam axis. Given that at the extreme energies of the QGP one can neglect the masses of the quarks and assume the matter to be approximately conformal, Bjorken flow predicts the energy density $\epsilon$ of the QGP to evolve as $\epsilon \propto \tau^{-4/3}$, where one has to make use of the conformal equation of state relating the pressure $P$ to the energy density via $P = \epsilon/3$.

The presumption of boost invariance that underlies Bjorken flow works well in the central rapidity region, whilst rotation invariance in the transverse plane is also a symmetry assuming the heavy nuclei collide head-on. However, the assumption of translation invariance in the transverse plane is quite drastic, as it does not allow for the possibility of a radial flow, and is thus certainly an oversimplification. In \cite{Gubser:2010ze}, Gubser proposed a generalization of Bjorken flow, referred commonly to as \emph{Gubser flow}, which does allow for a nontrivial radial flow profile for the QGP in the transverse plane. A key ingredient of Gubser flow is the assumption of conformal symmetry, which, as mentioned earlier, is well motivated at the energy scales involved. With the boost and rotation invariance along the beam axis kept intact, Gubser flow gets rid of exact translation invariance in the transverse plane, by replacing it with conformal invariance under a combination of translations and special conformal transformations. As it turns out, the equations of conformal relativistic hydrodynamics do admit a solution that has the above symmetries, with velocity and energy density profiles that now carry a radial dependence as well. 

From a geometric point of view, the ultrarelativistic nature of the hydrodynamic flow produced after a heavy-ion collision implies that the fluid is constrained to move on or near a null hypersurface. It must, therefore, be imbued with the underlying symmetries of this hypersurface. Carrollian symmetries are known to appear generically on null hypersurfaces.
The Carroll algebra arises as a contraction of the usual Poincar\'e algebra in the vanishing speed of light limit \emph{i.e.,} $c \to 0$ \cite{LevyLeblond, NDS}. One can construct a hydrodynamic description for fluids that are restricted to a Carroll manifold, dubbed \emph{Carroll hydrodynamics}, by starting from an ordinary relativistic fluid and carefully taking the $c \to 0$ limit \cite{Ciambelli:2018wre, Ciambelli:2018xat, Petkou:2022bmz, Freidel:2022bai}. This geometric picture for ultrarelativistic fluids was put on a firm footing in \cite{Bagchi:2023ysc}, where it was argued that Bjorken flow admits a dual description in terms of a Carroll fluid on a specific Carroll manifold. In particular, the phenomenological assumption of boost invariance of Bjorken flow follows from the underlying Carrollian symmetries in the dual description. Ref.~\cite{Bagchi:2023ysc} also provided the first concrete example of a physical realization of a Carroll fluid, which earlier had been thought to be only of theoretical interest in the context of e.g.~the flat version of the fluid-gravity correspondence. 

In the present work, we take another significant step forward in understanding ultrarelativistic fluids in terms of the Carroll framework. We will here go beyond the somewhat simplistic set-up of Bjorken flow and construct the dual Carrollian description for Gubser flow. It would provide us a geometric picture for this more general model for the evolution of QGP, which then paves the way to systematically compute rapidity-dependent corrections using the Carrollian perspective.

We achieve our objective by identifying a Carroll manifold conferred with a specific degenerate metric and kernel, such that the equations of conformal Carroll hydrodynamics on this manifold are the Gubser flow equations. As is the case with Bjorken flow \cite{Bagchi:2023ysc}, the phenomenological assumption of boost invariance that is put into the Gubser model using a special velocity profile now arises simply as a consequence of the underlying symmetries of the Carroll manifold. It is important to note that the duality we discuss also works beyond the perfect fluid approximation, when derivative corrections start playing a salient role in the dynamics of the fluid, under sensible scaling assumptions for the relativistic hydrodynamic data in the $c\to 0$ limit. Another important point that we make in our work is that the duality is not specific to the choice of a particular coordinate system and we exemplify this by constructing the duality for two different coordinate systems in the Gubser flow and two sets of Carrollian data which specify two different Carroll manifolds. 

Before we proceed, it is interesting to point out that Carroll symmetries arises in a variety of other places as well. For instance, conformal Carroll algebra is known to be isomorphic to the Bondi-van der Burg-Metzner-Sachs (BMS) algebra \cite{Bondi:1962px, Sachs:1962zza}, which dictates the symmetries of asymptotically flat spacetimes \cite{Bagchi:2010zz, Duval:2014uva}. Consequently, Carrollian conformal field theories (CFT) living on the null boundary of asymptotically flat spacetimes can provide the dual holographic description for gravitational dynamics in the bulk. This active field of research is referred to as the Carrollian approach to flat holography \cite{Bagchi:2012cy, Barnich:2012aw, Bagchi:2012xr, Barnich:2012xq, Bagchi:2014iea, Bagchi:2016bcd, Donnay:2022aba, Bagchi:2022emh, Donnay:2022wvx,Bagchi:2023fbj,Saha:2023hsl,Nguyen:2023vfz}. Apart from this, Carroll symmetries are relevant for certain condensed matter systems \cite{Bagchi:2022eui, Bidussi:2021nmp}, in cosmology \cite{deBoer:2021jej}, for string theory in the tensionless limit \cite{Bagchi:2013bga, Bagchi:2015nca, Bagchi:2020fpr}, as well as black holes \cite{Penna:2018gfx, Donnay:2019jiz, Freidel:2022vjq, Redondo-Yuste:2022czg, Bagchi:2022iqb}. Recent work on Carroll gravity and associated solutions appears in \cite{deBoer:2023fnj, Ecker:2023uwm}.

This paper is organized as follows. In section \ref{sec:Gubser}, we attempt to provide a self-contained discussion of Gubser flow and its associated symmetries. In section \ref{sec:Carroll}, we provide a brief overview of the construction of Carroll hydrodynamics as the $c \to 0$ limit of relativistic hydrodynamics, and discuss the conformal limit. Next, in section \ref{sec:mapping}, we establish the duality between Gubser flow and conformal Carroll hydrodynamics by providing an explicit map between the two, considering the fluid to be perfect. Subsequently, in section \ref{sec:viscous}, we argue that the duality is not limited only to perfect fluids, and exhibit its validity in the presence of first order hydrodynamic derivative corrections. We conclude the paper with a discussion and an outlook in section \ref{sec:discussion}.

\section{Heavy-ion collisions \& Gubser flow}
\label{sec:Gubser}
Describing the spacetime evolution of the hot and dense state of matter produced in heavy-ion collisions, the quark-gluon plasma (QGP), is a challenging task. The situation can be remedied to a certain extent by utilizing the symmetries of the problem. For instance, an important simplification occurs by realizing that the collision of two nuclei, moving towards each other at ultrarelativistic speeds and thereby highly Lorentz contracted, appears approximately the same in all centre-of-mass-like Lorentz frames, implying an emergent boost invariance for the QGP produced after the collision. Further, assuming the nuclear collision occurs approximately head-on, the evolution of the plasma must respect rotational invariance along the beam axis \emph{i.e.} in the two-dimensional plane transverse to the direction of collision. Making use of these symmetry considerations, along with the assumption of translation invariance in the transverse plane, a hydrodynamic model describing the spacetime evolution of the QGP was proposed by Bjorken in \cite{Bjorken:1982qr}. The model has met with considerable success, not least because exact solutions to viscous hydrodynamic equations are scarce. Modern experiments do indeed confirm the almost perfect fluid like nature of the QGP, and Bjorken's model has played a vital role in providing us an intuitive understanding of the same.

Despite its resounding success, the imposition of translation invariance on the transverse plane in Bjorken flow is perhaps too stringent and unrealistic. For one, demanding translation invariance does away with any radial dependence in hydrodynamic quantities like the energy density $\epsilon$, pressure $P$ and temperature $T$ of the plasma. On the other hand, in an actual heavy-ion collision, various hydrodynamic quantities will have a fall-off as a function of the radial distance $r$ from the beam axis. To rectify this, it is imperative to look for hydrodynamic models that do not insist on exact translation invariance in the transverse plane, and thereby allow for a radial dependence in hydrodynamic quantities.

A hydrodynamic model that achieves this objective was proposed by Gubser in \cite{Gubser:2010ze}. Gubser flow, as it has come to be known, assumes that the dynamics of the QGP at the extreme energy scales involved in heavy-ion collisions can be taken to be conformal. The model further relaxes the demand of exact translation invariance with conformal invariance of the flow under a combination of translations and special conformal transformations. As argued in \cite{Gubser:2010ze}, the conformal relativistic hydrodynamic equations now admit a more general solution, which still maintains boost and rotation invariance about the beam axis, while at the same time exhibiting nontrivial radial dependence for the four-velocity profile of the flow, as well as for the hydrodynamic quantities, as desired. Gubser flow is thus a more realistic and accurate model for high-energy nuclear collisions and the subsequent evolution of the QGP. As the focus of the present work is on Gubser flow and its connection with Carroll hydrodynamics, we now proceed to give a more mathematical description of the symmetries and equations governing the flow, which would be pertinent for establishing the duality with a conformal Carroll fluid in the following sections. 

Working in Cartesian coordinates, we choose the $z$-axis to be the beam axis, whilst $x, y$ denote the axes in the transverse plane, and $t$ denotes the time. For the sake of convenience, we can choose $t = 0$ as the moment of collision between the two nuclei, and $z=0$ being the location of collision. For the plasma produced after the collision, demanding invariance of the flow under a spacetime transformation $x^\mu \rightarrow x^\mu + \xi^\mu(x)$ amounts to imposing $\pounds_\xi u^\mu = 0$, where $u^\mu$ denotes the four-velocity of the fluid and $\pounds_\xi$ is the Lie derivative with respect to $\xi^\mu$. For instance, imposing invariance under translations and rotations in the $x, y$ plane along with boost invariance along the $z$-axis uniquely fixes the fluid velocity to be $u^\mu = (\gamma,0,0,\gamma v)$, with $v = z/t$ and the Lorentz factor $\gamma = 1/\sqrt{1-v^2}$.\footnote{The fluid velocity is normalized such that $u^\m u_\m = -1$.} This is indeed the velocity profile that underlies Bjorken's model of spacetime evolution of the QGP, and expectedly follows from the symmetry assumptions we imposed to begin with.

Gubser's model for the hydrodynamic description of QGP invokes invariance of the flow under boosts along and rotations about the beam axis. The salient feature of the model, however, is to replace translation invariance along the $x, y$ axes with conformal invariance under a combination of translations and special conformal transformations. The generators $\xi \equiv \xi^\mu \del_\mu$ for these transformations have the form
\bes
\label{Gubser_gens}
\begin{align}
\xi_1 &= \del_x + q^2 \left(2 x x^\m \del_\m - x^\m x_\m \del_x\right),\\
\xi_2 &= \del_y + q^2 \left(2 y x^\m \del_\m - x^\m x_\m \del_y\right).
\end{align}
\ees
Clearly, these are combinations of translations generated by $\del_x / \del_y$, with a special conformal transformation generated by $2b^\nu x_\nu x^\mu \del_\mu - x^\mu x_\mu b^\nu \del_\nu$, with $b^\nu = \d^\nu_x/\d^\nu_y$ for the two cases respectively. Here $q$ is a tunable parameter that carries the dimensions of inverse length. In particular, in the limit $q \to 0$, the generators in eq.~\eqref{Gubser_gens} reduce to ordinary translation generators. Interestingly, the generators $\xi_1, \xi_2$ along with the generator of rotations in the $x-y$ plane, $\xi_{\rm rot} = x\del_y -y \del_x$, form an $SO(3)_q$ subgroup\footnote{The subscript in $SO(3)_q$ is to signify the inherent dependence of the group generators on the parameter $q$.} of the full conformal group $SO(4,2)$. It is indeed straightforward to check that we have the algebra
\be
\label{so3_group}
\left[\xi_1, \xi_2\right] = - 4q^2 \xi_{\rm rot} ,\,\,\,  \left[\xi_1, \xi_{\rm rot}\right] = \xi_2, \,\,\, \left[\xi_2, \xi_{\rm rot}\right] = - \xi_1.
\ee
Further, the $SO(3)_q$ group above commutes with the $SO(1,1)$ subgroup of $SO(4,2)$ corresponding to boosts along the $z$-axis, generated by $\xi_{\rm boost} = z\del_t + t\del_z$.

\subsection{Gubser Flow in Milne Coordinates}

\begin{figure}
\centering
\includegraphics[width=8.8cm]{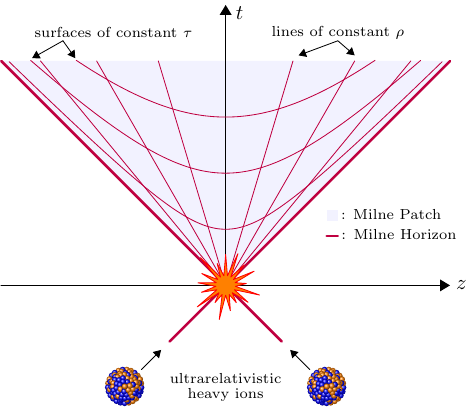}
\caption{A pictorial representation of the ultrarelativistic heavy-ion collision process. The heavy ions move at almost the speed of light. For simplicity, we assume that the collision occurs at time $t=0$ at position $z=0$. The forward lightcone of the collision event defines the Milne patch of Minkowski spacetime, covered by the coordinates $(\tau, \rho)$ - see eq.~\eqref{Milne_coords}.}
\label{Collision}
\end{figure}

A more convenient coordinate system for discussing the Gubser flow is provided by the Milne coordinates, denoted by $(\tau, \rho, r, \phi)$ - see fig.~\ref{Collision}. These are related to the Cartesian coordinates via
\be
\label{Milne_coords}
\begin{split}
\tau &= \sqrt{t^2 -z^2},\quad r = \sqrt{x^2 + y^2},\quad \phi = \tan^{-1}\left(\frac{y}{x}\right),\\
\rho &= \tanh^{-1}v = \tanh^{-1}\left(\frac{z}{t}\right) = \frac{1}{2} \log\left(\f{t+z}{t-z}\right).
\end{split}
\ee
Here $\tau$ is proper time and $\rho$ is the rapidity parameter, while $r, \phi$ are standard polar coordinates on the two-dimensional $x-y$ plane. 
The Minkowski metric in terms of the Milne coordinates has the form
\be
\label{Milne_metric}
ds^2 = - d\t^2 + \t^2 d\rho^2 + dr^2 + r^2 d\phi^2.
\ee
The generators eq.~\eqref{Gubser_gens} now take the form
\bes
\label{conf_generators}
\begin{align}
\xi_1 &= 2q^2 \t r \cos \phi \, \del_\t + [1+q^2 (\t^2 + r^2) ] \cos \phi \, \del_r \nonumber\\
&\quad-\f{1+q^2(\t^2-r^2)}{r} \sin\phi \, \del_\phi \, ,\\
\xi_2 &= 2q^2 \t r \sin \phi \, \del_\t + [1+q^2 (\t^2 + r^2) ] \sin \phi \, \del_r \nonumber\\
&\quad+\f{1+q^2(\t^2-r^2)}{r} \cos\phi \, \del_\phi \, ,
\end{align}
\ees
while the rotation and boost generators become $\xi_{\rm rot} = \del_\phi$ and $\xi_{\rm boost} = \del_\rho$. 

As mentioned earlier, the fluid velocity profile for Gubser flow follows by demanding invariance under rotations $\del_\phi$ and boosts $\del_\rho$. One further imposes a $\mathbb{Z}_2$ symmetry under $\rho \leftrightarrow - \rho$, reminiscent of the symmetry of the flow under $z\leftrightarrow -z$. This fixes the velocity profile to be\footnote{Notice that Gubser's model tacitly assumes the absence of an angular flow \emph{i.e.} $u^\phi = 0$. More generally, invariance under $\del_\phi, \del_\rho$ and the $\mathbb{Z}_2$ symmetry $\rho\leftrightarrow -\rho$ allows for a non-vanishing $u^\phi(\t, r)$.}
\be
\label{Gubser_velo}
u^\mu = \left(\cosh \kappa(\t, r), 0, \sinh \k(\t, r), 0\right).
\ee
It turns out that there does not exist any choice for $\k(\tau,r)$ such that $u^\m$ further respects invariance under the generators $\xi_1, \xi_2$. In \cite{Gubser:2010ze}, Gubser argues that it is apt to demand invariance of $u^\mu$ under $\xi_1, \xi_2$ only up to a conformal factor, reflecting the underlying conformal invariance of the setup.\footnote{The generators $\xi_a, a =1,2$ are conformal isometries for the background metric with the conformal factor $\half \nabla_\a \xi^\a_a$, \emph{i.e.,} $$\pounds_{\xi_a}g_{\m\n} = \half \left(\nabla_\a \xi^\a_a\right)g_{\mu\nu}.$$ It is thus justified to demand invariance of $u^\m$ under $\xi_a$ only up to a conformal factor proportional to $\nabla_\a \xi^\a_a$.} Thus, one only puts the requirement that $\pounds_{\xi_1}u^\mu \propto \left(\nabla_\nu \xi^\nu_1\right) u^\mu$, and similarly for $\xi_2$. This indeed allows for a unique solution for $\kappa(\t, r)$, given by
\be
\label{kappa_func}
\kappa(\t,r) = \tanh^{-1} \left[\f{2q^2 \t r}{1+q^2(\t^2 + r^2)}\right],
\ee
which leads to the fluid velocity eq.~\eqref{Gubser_velo} satisfying
\be
\label{Lie_der_x1x2}
\pounds_{\xi_1}u^\mu = - \f{1}{4} \left(\nabla_\nu \xi^\nu_1\right) u^\mu, \quad \pounds_{\xi_2}u^\mu = - \f{1}{4} \left(\nabla_\nu \xi_2^\nu\right) u^\mu.
\ee
To sum up, the fluid four velocity profile for Gubser flow takes the form of eq.~\eqref{Gubser_velo}, with $\kappa(\t, r)$ given by eq.~\eqref{kappa_func}. This velocity profile respects rotation and boost invariance along the beam axis, $\pounds_{\xi_{\rm rot}} u^\m = \pounds_{\xi_{\rm boost}} u^\m = 0$, but is only invariant up to a conformal factor under the generators $\xi_1, \xi_2$, as encapsulated by eq.~\eqref{Lie_der_x1x2}.\footnote{The $q\to 0$ limit reduces eq.~\eqref{Gubser_velo}, eq.~\eqref{kappa_func} to Bjorken flow, where the fluid happens to be at rest in Milne coordinates. Also, eq.~\eqref{Lie_der_x1x2} boils down to the fluid velocity respecting exact translation invariance.} Interestingly, the fluid velocity now has a nontrivial dependence on the radial coordinate $r$, apart from its dependence upon $\t$, and thus provides a more realistic description of the spacetime evolution of the QGP.

The next step is to consider the hydrodynamic equations themselves, which correspond to the conservation of the energy-momentum tensor $T^{\m\n}$ for the fluid, where
\be
\label{stress_ideal}
T^{\m\n} = \e u^\m u^\n + \f{\e}{3} \, \D^{\m\n},
\ee
with $\D_{\m\n} \equiv g_{\m\n} + u_\m u_\n$ being the projector orthogonal to $u^\m$. In writing eq.~\eqref{stress_ideal}, we have assumed the fluid to be ideal (see sec.~\ref{sec:viscous} for a discussion of viscous effects), and have also utilized the equation of state for a conformal fluid, which relates the pressure of the fluid to its energy density via $P = \e/3$. The hydrodynamic equations are
\be
\label{stress_cons}
\nabla_\m T^{\m\n} = 0.
\ee
Of these, the $\nu=\rho$ and $\nu=\phi$ components lead to the independence of the energy density from $\rho$ and $\phi$, respectively. This has been built into the system by the choice of the underlying symmetries and the associated velocity profile. Thus, the energy density can only depend upon $\t, r$. It is the $\nu=\t$ and $\nu=r$ components of eq.~\eqref{stress_cons} that indeed lead to nontrivial dynamical equations for $\e(\t,r)$, which can be combined suitably and expressed as
\bes
\label{Gubser_dynamics_1}
\begin{align}
\del_\t \e &= \f{4\e}{3}\left(\f{\cosh 2\k - 2}{\t} - \f{\sinh 2\k}{r}\right),\\
\del_r \e &= \f{4\e}{3} \left(\f{\cosh 2\k - 1}{r} - \f{\sinh 2\k}{\t}\right),
\end{align}
\ees
where $\k(\t,r)$ is given by eq.~\eqref{kappa_func}. These equations can be solved simultaneously and admit the solution
\be
\label{Gubser_energy_1}
\e(\t, r) = \f{\e_0}{\t^{4/3}} \f{(2q)^{8/3}}{[1+2q^2(\t^2+r^2)+q^4(\t^2-r^2)^2]^{4/3}} ,
\ee
where $\e_0$ is a constant of integration. As with the velocity profile for Gubser flow, we now have an interesting dependence on the radial coordinate $r$ in the energy density as well. Further, taking the $q\to 0$ limit with $\e_0 q^{8/3}$ held fixed in eq.~\eqref{Gubser_energy_1} leads to $\e \propto \t^{-4/3}$, which is the (radially independent) energy density for the Bjorken flow. 

\subsection{Gubser flow on de Sitter background}
\label{Gubser_dS}
Before we move on to the next section, it is useful to recast Gubser flow on another background, as originally proposed in \cite{Gubser:2010ui}. The new background corresponds to a product manifold comprising three-dimensional de Sitter (dS) spacetime times the real line $\mathbb{R}$. Starting from the flat background in Milne coordinates, one performs a Weyl rescaling of the metric eq.~\eqref{Milne_metric} via $ds^2 \to ds^2/\t^2$. This is followed by performing a coordinate transformation from $(\t, r)$ to $(\varsigma, \psi),$ where
\be
\label{flat_to_dS}
\begin{split}
\sinh \varsigma &= - \f{1-q^2(\t^2 - r^2)}{2q\t}\, ,\\
\tan \psi &= \f{2qr}{1+q^2(\t^2 - r^2)}\, .
\end{split}
\ee
The metric after these transformations becomes
\be
\label{dSxR}
ds^2 = - d\varsigma^2 + \cosh^2 \varsigma \left(d\psi^2 + \sin^2 \psi\, d\phi^2\right) + d\rho^2.
\ee
This is the metric on dS$_3 \times \mathbb{R}$, where $(\varsigma, \psi, \phi)$ are coordinates on the three-dimensional global de Sitter spacetime. The advantage of working on the dS$_3 \times \mathbb{R}$ background is that the $SO(3)_q$ conformal symmetry of flat spacetime now becomes an exact symmetry, as is evidenced by the presence of the two-sphere parametrized by $(\psi, \phi)$ in the metric eq.~\eqref{dSxR}. Algebraically, it is straightforward to check that the generators of $SO(3)_q$ conformal symmetry for the flat background, eq.~\eqref{conf_generators}, after the Weyl rescaling $ds^2 \to ds^2/\t^2$ and the coordinate transformation eq.~\eqref{flat_to_dS} become 
\bes
\begin{align}
\xi_1 &= 2q \cos\phi \, \del_\psi - 2q \cot\psi \sin\phi \, \del_\phi\, ,\\
\xi_2 &= 2q \sin\phi \, \del_\psi + 2q \cot\psi \cos\phi \, \del_\phi\, ,
\end{align}
\ees
which along with $\xi_{\rm rot} = \del_\phi$ are indeed the $SO(3)_q$ isometry generators for the two-sphere parametrized by $(\psi, \phi)$. Further, the dS$_3 \times \mathbb{R}$ background also has the $SO(1,1)$ boost invariance \emph{i.e.,} $\del_\rho$ is trivially an isometry for eq.~\eqref{dSxR}.
One can now construct a four-velocity profile for the fluid which respects the $SO(3)_q\times SO(1,1) \times \mathbb{Z}_2$ symmetry of the dS$_3 \times \mathbb{R}$ background. This is simply $u^\m = (1,0,0,0)$ \emph{i.e.,} the fluid is at rest on this background.

Interestingly, the $\n = \psi, \phi, \rho$ components of the hydrodynamic equations eq.~\eqref{stress_cons} with the energy-momentum tensor eq.~\eqref{stress_ideal} imply that the energy density is independent of $\psi, \phi$ and $\rho$, as expected from the underlying symmetries. The only nontrivial component of the hydrodynamic equations is the $\nu = \varsigma$ component, given by
\be
\label{Gubser_dynamics_2}
\del_\varsigma \e = - \f{8\e}{3} \tanh\varsigma\, ,
\ee
which has the solution 
\be
\label{Gubser_energy_2}
\e(\varsigma) = \f{\e_0}{(\cosh \varsigma)^{8/3}}\, . 
\ee

By undoing the coordinate transformation eq.~\eqref{flat_to_dS} along with the Weyl rescaling, it is straightforward to check that the energy density profile in eq.~\eqref{Gubser_energy_2} maps onto the profile in eq.~\eqref{Gubser_energy_1}. Therefore, in general, one may choose to perform the analysis for Gubser flow on the dS$_3\times\mathbb{R}$ background eq.~\eqref{dSxR}, where the velocity profile is elementary since the fluid is at rest, followed by undoing the Weyl rescaling and the coordinate transformation eq.~\eqref{flat_to_dS} to get back to flat background where the actual spacetime dynamics takes place. In what follows, we will establish a duality between the Gubser flow and conformal Carroll hydrodynamics, both for the plasma on the flat background, eq.~\eqref{Gubser_dynamics_1}, as well as on the dS$_3 \times \mathbb{R}$ background, eq.~\eqref{Gubser_dynamics_2}.

\section{Carroll hydrodynamics}
\label{sec:Carroll}
We now move on to provide an overview of Carroll hydrodynamics, setting up the notation and the key ideas that would then help us establish the duality with Gubser flow. As discussed in the Introduction, Carroll hydrodynamics arises in the vanishing speed of light $c\to 0$  limit of relativistic hydrodynamics \cite{Ciambelli:2018wre, Ciambelli:2018xat, Petkou:2022bmz, Freidel:2022bai}.\footnote{A more general class of Carroll fluids can be constructed by carefully accounting for the dynamics of the Goldstone mode associated with the spontaneous breaking of Carroll boost-invariance \cite{Armas:2023dcz}. For our purposes, however, the $c\to 0$ subclass suffices to establish the duality with Gubser flow.} Following \cite{Ciambelli:2018xat, Ciambelli:2018wre, Petkou:2022bmz}, it is convenient to work in the Papapetrou-Randers (PR) coordinates for imposing the $c\to 0$ limit, which makes manifest the Carroll covariance properties of various entities in a straight forward manner. A pseudo-Riemannian manifold in the PR parametrization \cite{Gibbons:2008zi, Ciambelli:2018xat} has the form 
\begin{align}
& ds^2 = g_{\mu\nu}dx^{\mu}dx^{\nu} = -c^2 (\Omega dt - b_i dx^i)^2 + a_{ij}dx^idx^j, \nonumber \\
& g^{00} = \frac{-1+c^2b^2}{\Omega^2}, \quad g^{0i} = g^{i0} = \frac{c}{\Omega} \, b^i, \quad g^{ij} = a^{ij}. \label{pseudo-Riem}
\end{align}
Here, $\Omega, b_i$ and $a_{ij}$ are functions of $(t, \textbf{x})$. Also, $b^2 \equiv b^i b_i$, where the index on $b_i$ can be raised using $a^{ij}$. 
On imposing the Carroll limit $c\to 0$ on eq.~\eqref{pseudo-Riem}, one arrives at a Carrollian manifold ${\cal C}$ described in terms of the degenerate metric $h_{\mu\nu}$ on ${\cal C}$ along with its kernel $k^\m$, expressed in coordinates $(t,\textbf{x})$ as
\begin{equation}\label{Carroll_metric}
d\ell^2 = h_{\mu\nu}dx^{\mu}dx^{\nu} = a_{ij}(t,\textbf{x})dx^i dx^j, \quad k = \frac{1}{\Omega(t,\textbf{x})}\,\partial_t,
\end{equation}
satisfying $h_{\mu\nu}k^{\nu} = 0$. In fact, the Carrollian manifold $\cal{C}$ is a fibre bundle, with time fibred over a base spatial manifold. Further, $b_i(t,\textbf{x})$ serves as the Ehresmann connection on ${\cal C}$, appearing naturally in the dual form to the kernel, $\vartheta$, via $\vartheta = \Omega dt - b_i dx^i$, such that $k^{\mu} \vartheta_{\mu} = 1$.
In particular, in this description, a flat Carroll manifold corresponds to
\begin{equation}
\Omega = 1, \quad a_{ij} = \delta_{ij}, \quad b_i = \text{constant},
\end{equation}
and its isometries generate the Carroll algebra, obtained in the $c\to 0$ limit of Poincar\'e algebra. 

On the the Carroll manifold eq.~\eqref{Carroll_metric}, one can construct several Carroll covariant quantities. For instance, the Carroll covariant temporal and spatial derivatives are
\begin{equation}
\label{Carr_dervs}
\hat{\partial}_t \equiv \frac{1}{\Omega}\,\partial_t\, , \quad \hat{\partial}_i \equiv \partial_i + \frac{b_i}{\Omega}\,\partial_t\, .
\end{equation}
Further, one can define temporal and spatial Levi-Civita-Carroll connections, respectively via
\begin{equation}
\hat{\gamma}_{ij} \equiv \frac{1}{2\Omega}\,\partial_t a_{ij}, \quad \hat{\gamma}^i_{jk} \equiv \frac{a^{il}}{2}\big(\hat{\partial}_j a_{kl} + \hat{\partial}_k a_{jl} - \hat{\partial}_l a_{jk} \big).
\end{equation}
Note that $\hat{\gamma}_{ij}$ is also a Carrollian tensor, and therefore its indices can be raised by using $a^{ij}$. The above definitions lead to temporal and spatial Levi-Civita-Carroll covariant derivatives, $\hat{\nabla}_t$ and $\hat{\nabla}_i$, respectively, which act on Carrollian objects via
\begin{equation}
\begin{split}
&\hat{\nabla}_t V^j = \hat{\partial}_t V^j + \hat{\gamma}^j_k V^k, \quad \hat{\nabla}_t V_j = \hat{\partial}_t V_j - \hat{\gamma}_j^k V_k,\\
&\hat{\nabla}_i V^j = \hat{\partial}_i V^j + \hat{\gamma}^j_{ik}V^k, \quad \hat{\nabla}_i V_j = \hat{\partial}_i V_j - \hat{\gamma}^k_{ij}V_k .
\end{split}
\end{equation}
Note that the connections above are metric compatible, \emph{i.e.} $\hat{\nabla}_t a_{jk} = 0, \hat{\nabla}_i a_{jk} = 0$. Further, one can also define a Carrollian expansion $\theta$ and a Carrollian acceleration $\varphi_i$ via
\begin{equation}
\label{Carr_exp_acc}
\theta \equiv \frac{1}{\Omega}\partial_t \log\sqrt{a} = \hat{\gamma}^i_i, \quad \varphi_i \equiv \frac{1}{\Omega}(\partial_t b_i + \partial_i\Omega),
\end{equation}
where $a = \det a_{ij}$.

With the basics of Carrollian geometry at hand, let us now proceed to summarize the construction of Carroll hydrodynamics as the $c\to 0$ limit of relativistic hydrodynamics. To begin with, we will consider a perfect fluid on the pseudo-Riemannian background eq.~\eqref{pseudo-Riem}, with the energy-momentum tensor
\begin{equation}
\label{stress_c_ideal}
T^{\mu\nu} = (\epsilon + P)\frac{u^{\mu}u^{\nu}}{c^2} + Pg^{\mu\nu}.
\end{equation}
Here $\epsilon, P$ are the energy density and pressure of the fluid, respectively, while $u^\mu$ is the fluid four-velocity, normalized such that $u^\mu u_\mu = -c^2$. A convenient parametrization for the fluid velocity turns out to be $u = \gamma\partial_t + \gamma v^i\partial_i$, where
\begin{equation}\label{PR_velocity}
\gamma = \frac{1+c^2\vec{b}\cdot\vec{\beta}}{\Omega\sqrt{1-c^2\beta^2}}\, , \quad v^i = \frac{c^2\Omega \beta^i}{1+c^2\vec{b}\cdot\vec{\beta}}\, ,
\end{equation}
with $\beta^i(t,\textbf{x})$ being a Carrollian vector field, whose index can therefore be lowered using $a_{ij}$, and $\beta^2 \equiv \beta^i\beta_i$, $\vec{b}\cdot\vec{\beta} \equiv b^i\beta_i = b_i \b^i$. To be explicit, the components of the fluid velocity in this parametrization take the form 
\begin{equation}
\label{velo_comps}
\begin{split}
&u^0 = \frac{c}{\O} \frac{1+c^2\vec{b}\cdot\vec{\b}}{\sqrt{1-c^2\b^2}}\, , \quad u^i = \frac{c^2 \b^i}{\sqrt{1-c^2\b^2}}\, ,\\
&u_0 = - \frac{c\,\O}{\sqrt{1-c^2\b^2}}\, , \quad\, u_i = \frac{c^2(b_i + \b_i)}{\sqrt{1-c^2\b^2}}\, . 
\end{split}
\end{equation}

Consider now the $c\to 0$ limit of the energy-momentum tensor eq.~\eqref{stress_c_ideal}. With the ansatz \cite{Ciambelli:2018wre}
\begin{equation}
\label{enery_pressure_expansion}
\epsilon = \varepsilon + {\cal O}(c^2)\, , \quad P = p + {\cal O}(c^2)
\end{equation}
for the leading behaviour of the energy density and pressure in the limit $c\to 0$, the components of the perfect fluid stress tensor become
\begin{equation}
\label{Carroll_stress_ideal}
\begin{split}
&T^0_{\,\,\, 0} = - \varepsilon + \mc{O}(c^2)\, , \quad T^i_{\,\, j} = p \, \d^i_{\,\, j} + \mc{O}(c^2)\, ,\\
&T^0_{\,\,\, i} = \frac{c}{\O} (\varepsilon + p) (b_i + \b_i) + \mc{O}(c^3)\, , \\
&T^i_{\,\,\,0} = - c\,\O(\varepsilon+p) \b^i + \mc{O}(c^3)\, .
\end{split}
\end{equation}
In obtaining eq.~\eqref{Carroll_stress_ideal}, we have made use of the $c\to 0$ behaviour of the fluid velocity components which follows from eq.~\eqref{velo_comps}, given by
\begin{equation}
\begin{split}
&u^0 = \frac{c}{\O} + \mc{O}(c^3)\, , \quad\quad u^i = c^2 \b^i + \mc{O}(c^4)\, ,\\
&u_0 = -c \, \O + \mc{O}(c^3)\, , \quad u_i = c^2 (b_i + \b_i) + \mc{O}(c^4)\, .
\end{split}
\end{equation}

Next, imposing the $c\to 0$ limit on the relativistic hydrodynamic equation $\nabla_{\mu} T^{\mu}_{\,\,\,\,\nu} = 0$ gives the hydrodynamic equations for a perfect Carroll fluid, 
\begin{subequations}\label{Carroll_eqns_ideal}
\begin{align}
\hat{\partial}_t \varepsilon &= -\theta(\varepsilon + p), \\
\hat{\partial}_i p &= -\varphi_i(\varepsilon + p) -\big(\hat{\partial}_t + \theta\big)[(\varepsilon + p)\beta_i],
\end{align}
\end{subequations}
where the Carrollian expansion $\theta$ and acceleration $\varphi_i$ have been defined in eq.~\eqref{Carr_exp_acc}, and the Carroll covariant temporal and spatial derivatives $\hat{\del}_t, \hat{\del}_i$ are defined in eq.~\eqref{Carr_dervs}.

\bigskip
\noindent \textit{Conformal Carroll fluid}
\medskip

For a conformal fluid, conformal invariance demands the vanishing of the trace of the energy-momentum tensor, which using eq.~\eqref{stress_c_ideal} furnishes an equation of state for the conformal fluid relating the energy density and pressure via $\epsilon = 3P$. Using the ansatz eq.~\eqref{enery_pressure_expansion}, this leads to the relation $\varepsilon = 3p$ at the leading order in $c\to 0$. Using this in the Carroll fluid equations \eqref{Carroll_eqns_ideal}, the hydrodynamic equations for a perfect conformal Carroll fluid become
\begin{subequations}
\label{conf_Carroll_eqns_ideal}
\begin{align}
\hat{\partial}_t \varepsilon &= -\frac{4}{3}\theta\varepsilon, \\
\hat{\partial}_i \varepsilon &= -4\varphi_i\varepsilon -4\big(\hat{\partial}_t + \theta\big)(\varepsilon\beta_i).
\end{align}
\end{subequations}

\section{From conformal Carroll hydrodynamics to Gubser flow}
\label{sec:mapping}
Equipped with an understanding of Gubser flow, section \ref{sec:Gubser}, and Carroll hydrodynamics, section \ref{sec:Carroll}, we now move on to establish a duality between the two, which forms the main result of the paper. To begin with, we present the duality when the fluid is ideal. In the next section, we will argue that the duality continues to hold true after including derivative corrections as well.

\subsection{Carroll fluid dual to Gubser flow in Milne}
We will begin by formulating our map in the coordinates we have introduced at the beginning, the Milne coordinates. We will show later that the mapping between Gubser flow and Carroll fluids continues to hold for other coordinate systems albeit with some different identifications. 

Given that Gubser flow assumes conformal invariance, we consider the conformal Carroll fluid equations \eqref{conf_Carroll_eqns_ideal} and adapt the coordinate chart on the Carroll manifold to Milne coordinates, \ie\ $(t,\textbf{x}) = (\tau,\rho,r,\phi)$. Then, by choosing the geometric data on this Carroll manifold such that
\begin{subequations}
\label{Gubser_Carroll_spacetime}
\begin{align}
& \Omega = \cosh\kappa, \\
& b_r = -\b_r+\sinh\kappa, \quad b_\rho = -\b_\rho, \quad b_\phi = -\b_\phi, \\
& a_{ij}dx^i dx^j = \tau^2 d\rho^2+\big[1+2q^2(\tau^2+r^2) \nonumber \\
& \hspace{18mm} +q^4(\tau^2-r^2)^2\big](dr^2 + r^2 d\phi^2),
\end{align}
\end{subequations}
with $\kappa$ defined in eq.~\eqref{kappa_func}, the hydrodynamic equations \eqref{conf_Carroll_eqns_ideal} for the conformal Carroll fluid become
\begin{subequations}
\label{conf_Carroll-Gubser_eqns_ideal}
\begin{align}
\partial_{\tau}\varepsilon &= \frac{4\varepsilon}{3}\left(\frac{\cosh 2\kappa - 2}{\tau} - \frac{\sinh 2\kappa}{r}\right), \label{carroll_first}\\
\partial_{r}\varepsilon &= \frac{4\varepsilon}{3}\left(\frac{\cosh 2\kappa - 1}{r} - \frac{\sinh 2\kappa}{\tau}\right), \label{carroll_second}\\
\partial_{\rho}\varepsilon &= 0, \quad \partial_{\phi}\varepsilon = 0. \label{carroll_third}
\end{align}
\end{subequations}
Equations \eqref{carroll_first} and \eqref{carroll_second} reproduce the dynamical equations \eqref{Gubser_dynamics_1} for Gubser flow. Further, eq.~\eqref{carroll_third} states that the energy density is independent of the rapidity and the angular coordinate $\phi$, implying boost invariance along the beam axis and rotational invariance in the transverse plane. Thus, we have derived the dynamical equations for Gubser flow along with the phenomenological assumptions of boost and rotation invariance from conformal Carroll hydrodynamics on a specific Carrollian manifold, establishing the duality between the two. 

\medskip

\subsection{Carroll fluid dual to Gubser flow on dS$_3\times \mathbb{R}$}

\medskip

The duality between Gubser flow and conformal Carroll hydrodynamics is not specific to the choice of Milne coordinates and holds more generally. We show this by establishing the duality in dS$_3\times\mathbb{R}$ coordinates as well. To do so, we now take the coordinate chart on the Carroll manifold adapted to the dS$_3\times\mathbb{R}$ coordinates, \ie\ $(t,\textbf{x}) = (\varsigma,\rho,\psi,\phi)$ and make the following choice for the geometric data on this manifold:
\begin{subequations}
\label{dS-Gubser_Carroll_spacetime}
\begin{align}
\Omega &= 1, \quad b_i = -\beta_i, \\
a_{ij}dx^idx^j &= \cosh^2 \varsigma \left(d\psi^2 + \sin^2 \psi\, d\phi^2\right) + d\rho^2.
\end{align}
\end{subequations}
On this background, the conformal Carroll hydrodynamic equations \eqref{conf_Carroll_eqns_ideal} reduce to
\begin{equation}
	\partial_{\varsigma}\varepsilon = -\frac{8\varepsilon}{3}\tanh\varsigma, \quad \partial_i\varepsilon = 0.
\end{equation}
The first equation reproduces the dynamical equation \eqref{Gubser_dynamics_2} for Gubser flow on dS$_3\times \mathbb{R}$ background. The second equation implies that the energy density is independent of the rapidity and the coordinates $\psi,\phi$ on the two-sphere, exhibiting invariance under the $SO(3)_q\times SO(1,1)\times\mathbb{Z}_2$ symmetries of the underlying dS$_3\times\mathbb{R}$ background.

\subsection{The $q\to 0$ limit and Bjorken flow}
\label{sec:qto0}
As was mentioned in section \ref{sec:Gubser}, Bjorken flow can be obtained as a limiting case from Gubser flow. In particular, the nontrivial radial evolution of the plasma in Gubser flow is parametrized by a nonzero value for $q$. Taking the limit $q\to 0$ leads to Bjorken flow, with exact translation invariance in the transverse plane. The dynamical equations \eqref{Gubser_dynamics_1} reduce in the $q\to 0$ limit to
\begin{equation}
	\partial_{\tau}\epsilon = -\frac{4\epsilon}{3\tau}, \quad \partial_r\epsilon = 0.
\end{equation}
The first equation is the conformal limit of the dynamical equation for Bjorken flow \cite{Bjorken:1982qr}, obtained by using the equation of state $\epsilon = 3P$. The second equation, combined with the independence of energy density from the coordinate $\phi$, reflects exact translation and rotation invariance of the flow in the transverse plane. 

\medskip

From the perspective of the dual conformal Carroll hydrodynamic description, on taking $q\rightarrow 0$ the geometric data of the dual Carroll manifold eq.~\eqref{Gubser_Carroll_spacetime} becomes
\begin{equation}
\label{Bjorken_Carroll_spacetime}
\Omega=1, \,\,\, b_i = -\beta_i, \,\,\, a_{ij}dx^idx^j = \tau^2 d\rho^2 + dr^2 + r^2 d\phi^2.
\end{equation}
On this Carroll manifold, the conformal Carroll hydrodynamic equations \eqref{conf_Carroll-Gubser_eqns_ideal} become
\begin{equation}
	\partial_{\tau}\varepsilon = -\frac{4\varepsilon}{3\tau}, \quad \partial_i\varepsilon = 0.
\end{equation}
The first equation reproduces the dynamical equation for  conformal Bjorken flow while the second equation reproduces the symmetries of Bjorken flow, \ie\ the boost invariance along the beam axis and the translational and rotational invariance in the transverse plane. 

As a matter of fact, the Carroll manifold eq.~\eqref{Bjorken_Carroll_spacetime} obtained in the $q\rightarrow 0$ limit is exactly the dual found in \cite{Bagchi:2023ysc} which originally established equivalence between Bjorken flow and Carroll hydrodynamics.

\section{Beyond perfect fluids}
\label{sec:viscous}
We now extend the duality between Gubser flow and conformal Carroll hydrodynamics discussed in the previous section to include derivative corrections beyond the perfect fluid form of the energy-momentum tensor. Including derivative corrections, one has
\begin{equation}
T^{\mu\nu} = T^{\mu\nu}_{\rm perf.} + T^{\m\n}_{(1)}+ T^{\m\n}_{(2)} + \cdots 
\end{equation}
Here $T^{\mu\nu}_{\rm perf.}$ is the perfect fluid energy-momentum tensor, eq.~\eqref{stress_c_ideal}, while the  $T^{\m\n}_{(i)}$ contain terms at the $i^{\rm th}$ order in derivatives of hydrodynamic variables, with $i=1,2,\ldots$. Thus $T^{\m\n}_{(1)}$ is the first order in derivatives correction to the perfect fluid energy-momentum tensor, $T^{\m\n}_{(2)}$ is the second order in derivatives correction, and so on. Our interest will be limited to considering terms only up to the first order in derivatives, while second order derivative corrections will be the subject of study in a future work.

To unambiguously write the derivative corrections, one needs to make the choice for a hydrodynamic frame, which essentially fixes the ambiguity that arises in the definition for hydrodynamic variables ones derivative corrections become non-negligible. For our discussion, we employ the Landau frame, which demands that the derivative corrections satisfy $T^{\m\n}_{(i)} u_\m = 0$. We also want the fluid to be conformal, \ie\ along with $\e=3P$, the derivative corrections must be traceless, $T^\m_{\,\,\,\m (i)} = 0$. These transverse and traceless conditions severely restrict the form of the derivative corrections, as we discuss below.

\subsection{First-order derivative corrections}
\label{sec:first-order}
At the first order in derivatives in the Landau frame, one has
\begin{equation}
T^{\m\n}_{(1)} = -\eta \sigma^{\mu\nu} - \zeta\Theta\Delta^{\mu\nu},
\end{equation}
where $\eta$, $\zeta$ are the shear and bulk viscosities respectively, $\Theta \equiv \nabla\cdot u$ is the divergence of fluid velocity, and $\sigma^{\mu\nu}$ is the shear tensor, given by
\begin{equation}
\sigma^{\mu\nu} = \Delta^{\mu\alpha}\Delta^{\nu\beta}\Big(\frac{\nabla_{\alpha}u_{\beta}+\nabla_{\beta}u_{\alpha}}{2}\Big)-\frac{1}{3}\Delta^{\mu\nu}\Theta.
\end{equation}
For the conformal case, the condition $T^\m_{\,\,\,\m(1)}=0$ leads to the vanishing of bulk viscosity, $\zeta=0$. One therefore has
\begin{equation}
\label{stress_firstorder}
T^{\mu\nu} = \epsilon u^{\mu}u^{\nu} + \frac{\epsilon}{3}\Delta^{\mu\nu} - \eta \sigma^{\mu\nu}.
\end{equation}

For a conformal fluid, the shear viscosity scales as $\eta = \eta_o \epsilon^{3/4}$, where $\eta_o$ is a dimensionless constant. 
Then, with the velocity profile eq.~\eqref{Gubser_velo}, the $\nu = \tau, r$ components of the hydrodynamic equations $\nabla_{\mu}T^{\mu\nu}=0$ for the stress tensor eq.~\eqref{stress_firstorder} give\footnote{The $\n = \rho, \phi$ components of the hydrodynamic equations still imply invariance of the flow under boosts and rotations, as encoded in the choice of the velocity profile.}
\bes
\label{Gubser_dynamics_firstorder}
\begin{align}
\del_\t \e &= \f{4\e}{3}\left(\f{\cosh 2\k - 2}{\t} - \f{\sinh 2\k}{r}\right) \nonumber \\
&\quad+ \f{2 \eta_o \e^{3/4}}{3 \, {\rm sech}^3\k}\left(
\f{1}{\t}-\f{\tanh\kappa}{r} \right)^2,\\
\del_r \e &= \f{4\e}{3} \left(\f{\cosh 2\k - 1}{r} - \f{\sinh 2\k}{\t}\right) \nonumber \\
&\quad- \f{2 \eta_o \e^{3/4}\sinh\kappa}{3 \, {\rm sech}^2\k}\left(
\f{1}{\t}-\f{\tanh\kappa}{r} \right)^2.
\end{align}
\ees
These are the viscous hydrodynamic equations governing the evolution of the QGP undergoing Gubser flow.\footnote{For large values of $\eta_o$, the equations \eqref{Gubser_dynamics_firstorder} may not admit a physically sensible solution; see \cite{Gubser:2010ze, Gubser:2010ui} for discussions on this issue. This will not play an important role for us, as our motive is to establish the duality with Carroll hydrodynamics at the level of the dynamical equations themselves.}

\bigskip

\noindent\emph{Carroll limit and the dual viscous conformal Carroll fluid}

\medskip

Let us now work out the hydrodynamic equations for a conformal Carroll fluid with viscous corrections. Reinstating the factors of $c$, the energy-momentum tensor for a relativistic viscous conformal fluid is given by
\begin{equation}
\label{stress_c_firstorder}
T^{\mu\nu} = (\e+P)\f{u^\m u^\n}{c^2} + Pg^{\m\n} -\eta \s^{\m\n}.
\end{equation}
The new element here is the presence of the shear tensor. Making use of the generic velocity parametrization of eq.~\eqref{PR_velocity}, and taking the $c\to 0$ limit gives
\be
\label{shear_behaviour}
\begin{split}
\s^0_{\,\,\,0} &= \mc{O}(c^2)\, , \quad\,\, \s^0_{\,\,\,i} = \f{c}{\O} (b^j+\b^j) \xi_{ij} + \mc{O}(c^3)\, , \\
\s^i_{\,\,0} &= - c\O\b_j\xi^{ij} + \mc{O}(c^3)\, , \quad \s^i_{\,\,j} = \xi^i_{\,\,j} + \mc{O}(c^2)\, ,
\end{split}
\ee
where $\xi_{ij}$ is the symmetric traceless Carrollian shear tensor, defined via
\begin{equation}
\xi_{ij} \equiv \frac{1}{2\Omega}\partial_t a_{ij}-\frac{\theta}{3}a_{ij}\, .
\end{equation}
Finally, to compute the Carroll hydrodynamic equations with viscous effects, \ie~the $c\to 0$ limit of $\N_\m T^\m_{\,\,\,\,\n} = 0$ with the energy-momentum tensor eq.~\eqref{stress_c_firstorder}, along with eqs.~\eqref{enery_pressure_expansion}, \eqref{Carroll_stress_ideal} and \eqref{shear_behaviour}, we make use of the ansatz
\begin{equation}
\label{eta_scaling}
\eta = \tilde{\eta} + {\cal O}(c^2).
\end{equation}
Then, the hydrodynamic equations for a viscous conformal Carroll fluid turn out to be
\begin{subequations}
\label{conf_Carroll_eqns_firstorder}
\begin{align}
\hat{\partial}_t \varepsilon &= -\frac{4}{3}\theta\varepsilon + \tilde{\eta} \xi^{ij}\xi_{ij}, \\
\hat{\partial}_i \varepsilon &= -4\varphi_i\varepsilon + 3\, \big(\hat\N_j + \varphi_j\big)\big[\tilde{\eta} \xi^j_{\,\,i}\big] \nonumber\\
&\quad\,- \big(\hat{\del}_t + \th\big)\big[4\varepsilon\b_i - 3\tilde{\eta} \b_j \xi^j_{\,\,i}\big].
\end{align}
\end{subequations}


We would now like to check whether the duality between Gubser flow and conformal Carroll hydrodynamics presented in section \ref{sec:mapping} continues to hold after the inclusion of first order derivative corrections. Specializing to Milne coordinates on the Carroll manifold, $(t,\textbf{x}) = (\tau,\rho,r,\phi)$, and choosing the geometric data to be eq.~\eqref{Gubser_Carroll_spacetime}, we find that the hydrodynamic equations \eqref{conf_Carroll_eqns_firstorder} for the viscous conformal Carroll fluid become
\bes
\label{conf_Carroll-Gubser_eqns_firstorder}
\begin{align}
\del_\t \varepsilon &= \f{4\varepsilon}{3}\left(\f{\cosh 2\k - 2}{\t} - \f{\sinh 2\k}{r}\right) \nonumber \\
&\quad+ \f{2 \, \eta_o \varepsilon^{3/4}}{3 \, {\rm sech}^3\k}\left(
\f{1}{\t}-\f{\tanh\kappa}{r} \right)^2,\\
\del_r \varepsilon &= \f{4\varepsilon}{3} \left(\f{\cosh 2\k - 1}{r} - \f{\sinh 2\k}{\t}\right) \nonumber \\
&\quad- \f{2 \eta_o \varepsilon^{3/4}\sinh\kappa}{3\, {\rm sech}^2\k}\left(
\f{1}{\t}-\f{\tanh\kappa}{r} \right)^2, \\
\del_{\rho} \varepsilon &= 0, \quad \del_{\phi}\varepsilon = 0 ,
\end{align}
\ees
where we have used $\tilde\eta = \eta_o \varepsilon^{3/4}$, which follows from the $c\to 0$ limit of $\eta = \eta_o \varepsilon^{3/4}$ combined with the scaling assumptions eqs.~\eqref{enery_pressure_expansion} and \eqref{eta_scaling}.  

The first two equations reproduce the dynamical equations \eqref{Gubser_dynamics_firstorder} of the viscous Gubser flow, while the third equation implies its boost and rotation invariance. Thus the duality between Gubser flow and conformal Carroll hydrodynamics continues to hold in the presence of first order hydrodynamic derivative corrections as well.

\bigskip

\noindent \emph{On {\rm{dS}}$_3\times\mathbb{R}$ background}

\medskip

As argued earlier as well, the proposed duality between Gubser flow and conformal Carroll hydrodynamics is not limited to any specific choice of coordinates. We observed this for the case of perfect fluids in section \ref{sec:mapping}, where the duality was shown to work for Gubser flow in both the Milne as well as the dS$_3\times \mathbb{R}$ coordinates. We would now like to verify the same with the inclusion of first order derivative corrections as well. 

On the dS$_3\times \mathbb{R}$ background, the hydrodynamic equation governing viscous Gubser flow is given by
\begin{equation}
\partial_{\varsigma}\epsilon = -\frac{8\epsilon}{3}\tanh\varsigma + \frac{2\eta_o\epsilon^{3/4}}{3}\tanh^2\varsigma.
\end{equation}

From the dual Carrollian perspective, the conformal Carroll hydrodynamic equations eq.~\eqref{conf_Carroll_eqns_firstorder} on the background eq.~\eqref{dS-Gubser_Carroll_spacetime} become
\begin{subequations}
\begin{align}
\partial_{\varsigma}\varepsilon &= -\frac{8\varepsilon}{3}\tanh\varsigma + \frac{2\eta_o\varepsilon^{3/4}}{3}\tanh^2\varsigma, \\
\partial_i\varepsilon &= 0,
\end{align}
\end{subequations}
thus, reproducing the dynamical equation and symmetries of viscous Gubser flow on the dS$_3\times\mathbb{R}$ background.

\section{Discussion and outlook}
\label{sec:discussion}
Heavy-ion collisions provide a complex laboratory for studying the properties of the strong force, which binds the quarks and gluons into nucleons, and the nucleons into nuclei. Thanks to the universality of hydrodynamics, we can understand the spacetime evolution of the QGP produced in such collisions in terms of analytically tractable models for ultrarelativistic fluids to a very good extent. The resulting picture that emerges is that of an almost perfect fluid with a very low shear viscosity to entropy density $\eta/s$ ratio, which undergoes rapid expansion, thermalization and hadronization into mesons and baryons that are eventually observed by the particle detectors.

In this paper, we have provided a dual geometric description for a prominent analytic model for heavy-ion collisions, namely Gubser flow. Assuming boost invariance along and rotation invariance about the beam axis, along with conformal invariance in the transverse plane, the model fixes a unique four-velocity profile for the QGP based purely on symmetries. Given an initial energy density profile $\epsilon(\tau_o, r)$, equations \eqref{Gubser_dynamics_firstorder} fix the spacetime evolution of the QGP, which can then be used to compare with the experimentally observed behaviour and extract the physical parameters of the plasma, such as its shear viscosity. In the dual description, the hydrodynamic equations governing Gubser flow arise naturally from the equations of Carroll hydrodynamics on a specific Carroll manifold. In particular, the key phenomenological assumption of boost invariance follows automatically from the geometric properties of this Carroll manifold. Put simply, the duality we propose geometrizes the otherwise phenomenological assumptions of Gubser flow in terms of Carrollian symmetries. This builds upon a similar dual description for another important model for the evolution of QGP, the Bjorken flow model, whose Carrollian dual was discussed in \cite{Bagchi:2023ysc}.  

The discovery of these dualities has now paved the way to systematically depart from the assumed symmetries and compute corrections to various analytic models for heavy-ion collisions. For instance, going beyond the leading order discussed in the present work, one can compute subleading terms arising in the $c\to 0$ limit in the hydrodynamic equations for a conformal Carroll fluid, and subsequently specialize to the Carroll manifold dual to Gubser flow. The subleading terms would then provide corrections to the Gubser flow equations that will encapsulate departures from exact boost invariance, as well as as other assumed symmetries. This, we expect, should then be able to provide a better analytic understanding as well as matching with the observed experimental data for heavy-ion collisions.

Apart from its relevance for the evolution of QGP, the duality between ultrarelativistic fluids and Carrollian dynamics raises many interesting questions and opens up a plethora of new directions for exploration. One such possibility is related to Carrollian dynamics arising on black hole horizons. As per the membrane paradigm for black holes, the event horizon of a black hole behaves like a fluid \cite{PhysRevD.18.3598, PhysRevD.33.915}. It was argued in \cite{Donnay:2019jiz} that the resulting hydrodynamic equations for this fluid are actually Carrollian in nature, something one would expect naturally given that event horizons are null hypersurfaces. With the advent of the duality between Carroll hydrodynamics and ultrarelativistic fluids, it now becomes an important question to understand how the dynamics of Bjorken/Gubser flow is related to the Damour and Raychaudhuri equations that govern the membrane paradigm. Another direction worth exploring is to understand the connection between generic null fluids and Carroll fluids. The concrete realization of Carroll fluids via the duality map in terms of Bjorken/Gubser flow models will provide a guiding light in exploring connections with generic null fluids. 

Finally, a word about holography. The fluid/gravity correspondence \cite{Bhattacharyya:2007vjd, Bhattacharyya:2008mz} has been extremely insightful in the context of AdS spacetimes and associated conformal fluids living on the boundary of AdS. The formulation of Carrollian fluids was initiated with the hope of constructing a version of the original fluid/gravity correspondence which would be of use for asymptotically flat spacetimes and offer hints into flat holography. Flat spacetimes can be thought of as a very high energy sector of AdS, where one is deep inside the AdS bulk and probing very small length scales compared to the radius of AdS, so that the curvature of AdS becomes imperceptible. Our construction of explicit examples of a duality between Carroll hydrodynamics and highly boosted ultrarelativistic fluids goes to show that the Carrollian sector of a relativistic fluid is indeed the very high energy sector and in keeping with the intuition from holography. The message is that fluid-gravity in the very highly boosted regime would become a duality between gravitational dynamics in asymptotically flat spacetimes and Carrollian fluids living on the null boundary. 

There have been studies of boost-invariant flows in the context of AdS/CFT, see \cite{Janik:2005zt, Janik:2006gp, Janik:2006ft, Grumiller:2008va, Albacete:2008vs, Heller:2008mb, Kinoshita:2008dq, Beuf:2009cx, Chesler:2009cy, Taliotis:2010pi} for early works. Our analysis in \cite{Bagchi:2023ysc} and in the current paper seems to suggest that these holographic studies are better suited to asymptotically flat spacetimes, given that there is an explicit map to Carrollian fluids. We wish to understand this rather intriguing point, as well as the several other questions mentioned above, in more detail in future work. 
\vspace{-6mm}
\acknowledgments
\vspace{-2mm}
We would like to thank Jay Armas, Daniel Grumiller, Emil Have, Paul Romatschke and Laurence Yaffe for discussions. AS would like to acknowledge the warm hospitality received from the Korea Institute for Advanced Study, Seoul, through their visitor program, as well as the Nordic Institute for Theoretical Physics (Nordita), Stockholm, during the workshop \emph{Hydrodynamics at All Scales,} and from the Aristotle University of Thessaloniki, Greece, during the \emph{3\textsuperscript{rd} Carroll Workshop}. 

AB is partially supported by a Swarnajayanti Fellowship from the Science and Engineering Research Board (SERB) under grant SB/SJF/2019-20/08 and also by the SERB grant CRG/2020/002035. The work of KK is partially supported by the grant SB/SJF/2019-20/08. The work of TM is supported by the grant SB/SJF/2019-20/08. The work of AS is supported by the European Research Council (ERC) under the European Union’s Horizon 2020 research and innovation programme (grant agreement no.~758759). 


\bibliography{Gub_Refs} 

\begin{thebibliography}{56}%
\makeatletter
\providecommand \@ifxundefined [1]{%
 \@ifx{#1\undefined}
}%
\providecommand \@ifnum [1]{%
 \ifnum #1\expandafter \@firstoftwo
 \else \expandafter \@secondoftwo
 \fi
}%
\providecommand \@ifx [1]{%
 \ifx #1\expandafter \@firstoftwo
 \else \expandafter \@secondoftwo
 \fi
}%
\providecommand \natexlab [1]{#1}%
\providecommand \enquote  [1]{``#1''}%
\providecommand \bibnamefont  [1]{#1}%
\providecommand \bibfnamefont [1]{#1}%
\providecommand \citenamefont [1]{#1}%
\providecommand \href@noop [0]{\@secondoftwo}%
\providecommand \href [0]{\begingroup \@sanitize@url \@href}%
\providecommand \@href[1]{\@@startlink{#1}\@@href}%
\providecommand \@@href[1]{\endgroup#1\@@endlink}%
\providecommand \@sanitize@url [0]{\catcode `\\12\catcode `\$12\catcode
  `\&12\catcode `\#12\catcode `\^12\catcode `\_12\catcode `\%12\relax}%
\providecommand \@@startlink[1]{}%
\providecommand \@@endlink[0]{}%
\providecommand \url  [0]{\begingroup\@sanitize@url \@url }%
\providecommand \@url [1]{\endgroup\@href {#1}{\urlprefix }}%
\providecommand \urlprefix  [0]{URL }%
\providecommand \Eprint [0]{\href }%
\providecommand \doibase [0]{https://doi.org/}%
\providecommand \selectlanguage [0]{\@gobble}%
\providecommand \bibinfo  [0]{\@secondoftwo}%
\providecommand \bibfield  [0]{\@secondoftwo}%
\providecommand \translation [1]{[#1]}%
\providecommand \BibitemOpen [0]{}%
\providecommand \bibitemStop [0]{}%
\providecommand \bibitemNoStop [0]{.\EOS\space}%
\providecommand \EOS [0]{\spacefactor3000\relax}%
\providecommand \BibitemShut  [1]{\csname bibitem#1\endcsname}%
\let\auto@bib@innerbib\@empty
\bibitem [{\citenamefont {Gale}\ \emph {et~al.}(2013)\citenamefont {Gale},
  \citenamefont {Jeon},\ and\ \citenamefont {Schenke}}]{Gale:2013da}%
  \BibitemOpen
  \bibfield  {author} {\bibinfo {author} {\bibfnamefont {C.}~\bibnamefont
  {Gale}}, \bibinfo {author} {\bibfnamefont {S.}~\bibnamefont {Jeon}},\ and\
  \bibinfo {author} {\bibfnamefont {B.}~\bibnamefont {Schenke}},\ }\bibfield
  {title} {\bibinfo {title} {{Hydrodynamic Modeling of Heavy-Ion Collisions}},\
  }\href {https://doi.org/10.1142/S0217751X13400113} {\bibfield  {journal}
  {\bibinfo  {journal} {Int. J. Mod. Phys. A}\ }\textbf {\bibinfo {volume}
  {28}},\ \bibinfo {pages} {1340011} (\bibinfo {year} {2013})},\ \Eprint
  {https://arxiv.org/abs/1301.5893} {arXiv:1301.5893 [nucl-th]} \BibitemShut
  {NoStop}%
\bibitem [{\citenamefont {Bjorken}(1983)}]{Bjorken:1982qr}%
  \BibitemOpen
  \bibfield  {author} {\bibinfo {author} {\bibfnamefont {J.~D.}\ \bibnamefont
  {Bjorken}},\ }\bibfield  {title} {\bibinfo {title} {{Highly Relativistic
  Nucleus-Nucleus Collisions: The Central Rapidity Region}},\ }\href
  {https://doi.org/10.1103/PhysRevD.27.140} {\bibfield  {journal} {\bibinfo
  {journal} {Phys. Rev. D}\ }\textbf {\bibinfo {volume} {27}},\ \bibinfo
  {pages} {140} (\bibinfo {year} {1983})}\BibitemShut {NoStop}%
\bibitem [{\citenamefont {Gubser}(2010)}]{Gubser:2010ze}%
  \BibitemOpen
  \bibfield  {author} {\bibinfo {author} {\bibfnamefont {S.~S.}\ \bibnamefont
  {Gubser}},\ }\bibfield  {title} {\bibinfo {title} {{Symmetry constraints on
  generalizations of Bjorken flow}},\ }\href
  {https://doi.org/10.1103/PhysRevD.82.085027} {\bibfield  {journal} {\bibinfo
  {journal} {Phys. Rev. D}\ }\textbf {\bibinfo {volume} {82}},\ \bibinfo
  {pages} {085027} (\bibinfo {year} {2010})},\ \Eprint
  {https://arxiv.org/abs/1006.0006} {arXiv:1006.0006 [hep-th]} \BibitemShut
  {NoStop}%
\bibitem [{\citenamefont {Lévy-Leblond}(1965)}]{LevyLeblond}%
  \BibitemOpen
  \bibfield  {author} {\bibinfo {author} {\bibfnamefont {J.-M.}\ \bibnamefont
  {Lévy-Leblond}},\ }\bibfield  {title} {\bibinfo {title} {Une nouvelle limite
  non-relativiste du groupe de poincaré},\ }\href {http://eudml.org/doc/75509}
  {\bibfield  {journal} {\bibinfo  {journal} {Annales de l'I.H.P. Physique
  Théorique}\ }\textbf {\bibinfo {volume} {3}},\ \bibinfo {pages} {1}
  (\bibinfo {year} {1965})}\BibitemShut {NoStop}%
\bibitem [{\citenamefont {Sen~Gupta}(1966)}]{NDS}%
  \BibitemOpen
  \bibfield  {author} {\bibinfo {author} {\bibfnamefont {N.~D.}\ \bibnamefont
  {Sen~Gupta}},\ }\bibfield  {title} {\bibinfo {title} {{On an Analogue of the
  Galileo Group}},\ }\href {https://doi.org/10.1007/BF02740871} {\bibfield
  {journal} {\bibinfo  {journal} {Il Nuovo Cimento A}\ }\textbf {\bibinfo
  {volume} {44}},\ \bibinfo {pages} {512–517} (\bibinfo {year}
  {1966})}\BibitemShut {NoStop}%
\bibitem [{\citenamefont {Ciambelli}\ \emph
  {et~al.}(2018{\natexlab{a}})\citenamefont {Ciambelli}, \citenamefont
  {Marteau}, \citenamefont {Petkou}, \citenamefont {Petropoulos},\ and\
  \citenamefont {Siampos}}]{Ciambelli:2018wre}%
  \BibitemOpen
  \bibfield  {author} {\bibinfo {author} {\bibfnamefont {L.}~\bibnamefont
  {Ciambelli}}, \bibinfo {author} {\bibfnamefont {C.}~\bibnamefont {Marteau}},
  \bibinfo {author} {\bibfnamefont {A.~C.}\ \bibnamefont {Petkou}}, \bibinfo
  {author} {\bibfnamefont {P.~M.}\ \bibnamefont {Petropoulos}},\ and\ \bibinfo
  {author} {\bibfnamefont {K.}~\bibnamefont {Siampos}},\ }\bibfield  {title}
  {\bibinfo {title} {{Flat holography and Carrollian fluids}},\ }\href
  {https://doi.org/10.1007/JHEP07(2018)165} {\bibfield  {journal} {\bibinfo
  {journal} {JHEP}\ }\textbf {\bibinfo {volume} {07}},\ \bibinfo {pages}
  {165}},\ \Eprint {https://arxiv.org/abs/1802.06809} {arXiv:1802.06809
  [hep-th]} \BibitemShut {NoStop}%
\bibitem [{\citenamefont {Ciambelli}\ \emph
  {et~al.}(2018{\natexlab{b}})\citenamefont {Ciambelli}, \citenamefont
  {Marteau}, \citenamefont {Petkou}, \citenamefont {Petropoulos},\ and\
  \citenamefont {Siampos}}]{Ciambelli:2018xat}%
  \BibitemOpen
  \bibfield  {author} {\bibinfo {author} {\bibfnamefont {L.}~\bibnamefont
  {Ciambelli}}, \bibinfo {author} {\bibfnamefont {C.}~\bibnamefont {Marteau}},
  \bibinfo {author} {\bibfnamefont {A.~C.}\ \bibnamefont {Petkou}}, \bibinfo
  {author} {\bibfnamefont {P.~M.}\ \bibnamefont {Petropoulos}},\ and\ \bibinfo
  {author} {\bibfnamefont {K.}~\bibnamefont {Siampos}},\ }\bibfield  {title}
  {\bibinfo {title} {{Covariant Galilean versus Carrollian hydrodynamics from
  relativistic fluids}},\ }\href {https://doi.org/10.1088/1361-6382/aacf1a}
  {\bibfield  {journal} {\bibinfo  {journal} {Class. Quant. Grav.}\ }\textbf
  {\bibinfo {volume} {35}},\ \bibinfo {pages} {165001} (\bibinfo {year}
  {2018}{\natexlab{b}})},\ \Eprint {https://arxiv.org/abs/1802.05286}
  {arXiv:1802.05286 [hep-th]} \BibitemShut {NoStop}%
\bibitem [{\citenamefont {Petkou}\ \emph {et~al.}(2022)\citenamefont {Petkou},
  \citenamefont {Petropoulos}, \citenamefont {Betancour},\ and\ \citenamefont
  {Siampos}}]{Petkou:2022bmz}%
  \BibitemOpen
  \bibfield  {author} {\bibinfo {author} {\bibfnamefont {A.~C.}\ \bibnamefont
  {Petkou}}, \bibinfo {author} {\bibfnamefont {P.~M.}\ \bibnamefont
  {Petropoulos}}, \bibinfo {author} {\bibfnamefont {D.~R.}\ \bibnamefont
  {Betancour}},\ and\ \bibinfo {author} {\bibfnamefont {K.}~\bibnamefont
  {Siampos}},\ }\bibfield  {title} {\bibinfo {title} {{Relativistic fluids,
  hydrodynamic frames and their Galilean versus Carrollian avatars}},\ }\href
  {https://doi.org/10.1007/JHEP09(2022)162} {\bibfield  {journal} {\bibinfo
  {journal} {JHEP}\ }\textbf {\bibinfo {volume} {09}},\ \bibinfo {pages}
  {162}},\ \Eprint {https://arxiv.org/abs/2205.09142} {arXiv:2205.09142
  [hep-th]} \BibitemShut {NoStop}%
\bibitem [{\citenamefont {Freidel}\ and\ \citenamefont
  {Jai-akson}(2023)}]{Freidel:2022bai}%
  \BibitemOpen
  \bibfield  {author} {\bibinfo {author} {\bibfnamefont {L.}~\bibnamefont
  {Freidel}}\ and\ \bibinfo {author} {\bibfnamefont {P.}~\bibnamefont
  {Jai-akson}},\ }\bibfield  {title} {\bibinfo {title} {{Carrollian
  hydrodynamics from symmetries}},\ }\href
  {https://doi.org/10.1088/1361-6382/acb194} {\bibfield  {journal} {\bibinfo
  {journal} {Class. Quant. Grav.}\ }\textbf {\bibinfo {volume} {40}},\ \bibinfo
  {pages} {055009} (\bibinfo {year} {2023})},\ \Eprint
  {https://arxiv.org/abs/2209.03328} {arXiv:2209.03328 [hep-th]} \BibitemShut
  {NoStop}%
\bibitem [{\citenamefont {Bagchi}\ \emph
  {et~al.}(2023{\natexlab{a}})\citenamefont {Bagchi}, \citenamefont {Kolekar},\
  and\ \citenamefont {Shukla}}]{Bagchi:2023ysc}%
  \BibitemOpen
  \bibfield  {author} {\bibinfo {author} {\bibfnamefont {A.}~\bibnamefont
  {Bagchi}}, \bibinfo {author} {\bibfnamefont {K.~S.}\ \bibnamefont
  {Kolekar}},\ and\ \bibinfo {author} {\bibfnamefont {A.}~\bibnamefont
  {Shukla}},\ }\bibfield  {title} {\bibinfo {title} {{Carrollian Origins of
  Bjorken Flow}},\ }\href {https://doi.org/10.1103/PhysRevLett.130.241601}
  {\bibfield  {journal} {\bibinfo  {journal} {Phys. Rev. Lett.}\ }\textbf
  {\bibinfo {volume} {130}},\ \bibinfo {pages} {241601} (\bibinfo {year}
  {2023}{\natexlab{a}})},\ \Eprint {https://arxiv.org/abs/2302.03053}
  {arXiv:2302.03053 [hep-th]} \BibitemShut {NoStop}%
\bibitem [{\citenamefont {Bondi}\ \emph {et~al.}(1962)\citenamefont {Bondi},
  \citenamefont {van~der Burg},\ and\ \citenamefont {Metzner}}]{Bondi:1962px}%
  \BibitemOpen
  \bibfield  {author} {\bibinfo {author} {\bibfnamefont {H.}~\bibnamefont
  {Bondi}}, \bibinfo {author} {\bibfnamefont {M.~G.~J.}\ \bibnamefont {van~der
  Burg}},\ and\ \bibinfo {author} {\bibfnamefont {A.~W.~K.}\ \bibnamefont
  {Metzner}},\ }\bibfield  {title} {\bibinfo {title} {{Gravitational waves in
  general relativity. 7. Waves from axisymmetric isolated systems}},\ }\href
  {https://doi.org/10.1098/rspa.1962.0161} {\bibfield  {journal} {\bibinfo
  {journal} {Proc. Roy. Soc. Lond. A}\ }\textbf {\bibinfo {volume} {269}},\
  \bibinfo {pages} {21} (\bibinfo {year} {1962})}\BibitemShut {NoStop}%
\bibitem [{\citenamefont {Sachs}(1962)}]{Sachs:1962zza}%
  \BibitemOpen
  \bibfield  {author} {\bibinfo {author} {\bibfnamefont {R.}~\bibnamefont
  {Sachs}},\ }\bibfield  {title} {\bibinfo {title} {{Asymptotic symmetries in
  gravitational theory}},\ }\href {https://doi.org/10.1103/PhysRev.128.2851}
  {\bibfield  {journal} {\bibinfo  {journal} {Phys. Rev.}\ }\textbf {\bibinfo
  {volume} {128}},\ \bibinfo {pages} {2851} (\bibinfo {year}
  {1962})}\BibitemShut {NoStop}%
\bibitem [{\citenamefont {Bagchi}(2010)}]{Bagchi:2010zz}%
  \BibitemOpen
  \bibfield  {author} {\bibinfo {author} {\bibfnamefont {A.}~\bibnamefont
  {Bagchi}},\ }\bibfield  {title} {\bibinfo {title} {{Correspondence between
  Asymptotically Flat Spacetimes and Nonrelativistic Conformal Field
  Theories}},\ }\href {https://doi.org/10.1103/PhysRevLett.105.171601}
  {\bibfield  {journal} {\bibinfo  {journal} {Phys. Rev. Lett.}\ }\textbf
  {\bibinfo {volume} {105}},\ \bibinfo {pages} {171601} (\bibinfo {year}
  {2010})},\ \Eprint {https://arxiv.org/abs/1006.3354} {arXiv:1006.3354
  [hep-th]} \BibitemShut {NoStop}%
\bibitem [{\citenamefont {Duval}\ \emph {et~al.}(2014)\citenamefont {Duval},
  \citenamefont {Gibbons},\ and\ \citenamefont {Horvathy}}]{Duval:2014uva}%
  \BibitemOpen
  \bibfield  {author} {\bibinfo {author} {\bibfnamefont {C.}~\bibnamefont
  {Duval}}, \bibinfo {author} {\bibfnamefont {G.~W.}\ \bibnamefont {Gibbons}},\
  and\ \bibinfo {author} {\bibfnamefont {P.~A.}\ \bibnamefont {Horvathy}},\
  }\bibfield  {title} {\bibinfo {title} {{Conformal Carroll groups and BMS
  symmetry}},\ }\href {https://doi.org/10.1088/0264-9381/31/9/092001}
  {\bibfield  {journal} {\bibinfo  {journal} {Class. Quant. Grav.}\ }\textbf
  {\bibinfo {volume} {31}},\ \bibinfo {pages} {092001} (\bibinfo {year}
  {2014})},\ \Eprint {https://arxiv.org/abs/1402.5894} {arXiv:1402.5894
  [gr-qc]} \BibitemShut {NoStop}%
\bibitem [{\citenamefont {Bagchi}\ and\ \citenamefont
  {Fareghbal}(2012)}]{Bagchi:2012cy}%
  \BibitemOpen
  \bibfield  {author} {\bibinfo {author} {\bibfnamefont {A.}~\bibnamefont
  {Bagchi}}\ and\ \bibinfo {author} {\bibfnamefont {R.}~\bibnamefont
  {Fareghbal}},\ }\bibfield  {title} {\bibinfo {title} {{BMS/GCA Redux: Towards
  Flatspace Holography from Non-Relativistic Symmetries}},\ }\href
  {https://doi.org/10.1007/JHEP10(2012)092} {\bibfield  {journal} {\bibinfo
  {journal} {JHEP}\ }\textbf {\bibinfo {volume} {10}},\ \bibinfo {pages}
  {092}},\ \Eprint {https://arxiv.org/abs/1203.5795} {arXiv:1203.5795 [hep-th]}
  \BibitemShut {NoStop}%
\bibitem [{\citenamefont {Barnich}\ \emph {et~al.}(2012)\citenamefont
  {Barnich}, \citenamefont {Gomberoff},\ and\ \citenamefont
  {Gonzalez}}]{Barnich:2012aw}%
  \BibitemOpen
  \bibfield  {author} {\bibinfo {author} {\bibfnamefont {G.}~\bibnamefont
  {Barnich}}, \bibinfo {author} {\bibfnamefont {A.}~\bibnamefont {Gomberoff}},\
  and\ \bibinfo {author} {\bibfnamefont {H.~A.}\ \bibnamefont {Gonzalez}},\
  }\bibfield  {title} {\bibinfo {title} {{The Flat limit of three dimensional
  asymptotically anti-de Sitter spacetimes}},\ }\href
  {https://doi.org/10.1103/PhysRevD.86.024020} {\bibfield  {journal} {\bibinfo
  {journal} {Phys. Rev. D}\ }\textbf {\bibinfo {volume} {86}},\ \bibinfo
  {pages} {024020} (\bibinfo {year} {2012})},\ \Eprint
  {https://arxiv.org/abs/1204.3288} {arXiv:1204.3288 [gr-qc]} \BibitemShut
  {NoStop}%
\bibitem [{\citenamefont {Bagchi}\ \emph {et~al.}(2013)\citenamefont {Bagchi},
  \citenamefont {Detournay}, \citenamefont {Fareghbal},\ and\ \citenamefont
  {Sim\'on}}]{Bagchi:2012xr}%
  \BibitemOpen
  \bibfield  {author} {\bibinfo {author} {\bibfnamefont {A.}~\bibnamefont
  {Bagchi}}, \bibinfo {author} {\bibfnamefont {S.}~\bibnamefont {Detournay}},
  \bibinfo {author} {\bibfnamefont {R.}~\bibnamefont {Fareghbal}},\ and\
  \bibinfo {author} {\bibfnamefont {J.}~\bibnamefont {Sim\'on}},\ }\bibfield
  {title} {\bibinfo {title} {{Holography of 3D Flat Cosmological Horizons}},\
  }\href {https://doi.org/10.1103/PhysRevLett.110.141302} {\bibfield  {journal}
  {\bibinfo  {journal} {Phys. Rev. Lett.}\ }\textbf {\bibinfo {volume} {110}},\
  \bibinfo {pages} {141302} (\bibinfo {year} {2013})},\ \Eprint
  {https://arxiv.org/abs/1208.4372} {arXiv:1208.4372 [hep-th]} \BibitemShut
  {NoStop}%
\bibitem [{\citenamefont {Barnich}(2012)}]{Barnich:2012xq}%
  \BibitemOpen
  \bibfield  {author} {\bibinfo {author} {\bibfnamefont {G.}~\bibnamefont
  {Barnich}},\ }\bibfield  {title} {\bibinfo {title} {{Entropy of
  three-dimensional asymptotically flat cosmological solutions}},\ }\href
  {https://doi.org/10.1007/JHEP10(2012)095} {\bibfield  {journal} {\bibinfo
  {journal} {JHEP}\ }\textbf {\bibinfo {volume} {10}},\ \bibinfo {pages}
  {095}},\ \Eprint {https://arxiv.org/abs/1208.4371} {arXiv:1208.4371 [hep-th]}
  \BibitemShut {NoStop}%
\bibitem [{\citenamefont {Bagchi}\ \emph {et~al.}(2015)\citenamefont {Bagchi},
  \citenamefont {Basu}, \citenamefont {Grumiller},\ and\ \citenamefont
  {Riegler}}]{Bagchi:2014iea}%
  \BibitemOpen
  \bibfield  {author} {\bibinfo {author} {\bibfnamefont {A.}~\bibnamefont
  {Bagchi}}, \bibinfo {author} {\bibfnamefont {R.}~\bibnamefont {Basu}},
  \bibinfo {author} {\bibfnamefont {D.}~\bibnamefont {Grumiller}},\ and\
  \bibinfo {author} {\bibfnamefont {M.}~\bibnamefont {Riegler}},\ }\bibfield
  {title} {\bibinfo {title} {{Entanglement entropy in Galilean conformal field
  theories and flat holography}},\ }\href
  {https://doi.org/10.1103/PhysRevLett.114.111602} {\bibfield  {journal}
  {\bibinfo  {journal} {Phys. Rev. Lett.}\ }\textbf {\bibinfo {volume} {114}},\
  \bibinfo {pages} {111602} (\bibinfo {year} {2015})},\ \Eprint
  {https://arxiv.org/abs/1410.4089} {arXiv:1410.4089 [hep-th]} \BibitemShut
  {NoStop}%
\bibitem [{\citenamefont {Bagchi}\ \emph
  {et~al.}(2016{\natexlab{a}})\citenamefont {Bagchi}, \citenamefont {Basu},
  \citenamefont {Kakkar},\ and\ \citenamefont {Mehra}}]{Bagchi:2016bcd}%
  \BibitemOpen
  \bibfield  {author} {\bibinfo {author} {\bibfnamefont {A.}~\bibnamefont
  {Bagchi}}, \bibinfo {author} {\bibfnamefont {R.}~\bibnamefont {Basu}},
  \bibinfo {author} {\bibfnamefont {A.}~\bibnamefont {Kakkar}},\ and\ \bibinfo
  {author} {\bibfnamefont {A.}~\bibnamefont {Mehra}},\ }\bibfield  {title}
  {\bibinfo {title} {{Flat Holography: Aspects of the dual field theory}},\
  }\href {https://doi.org/10.1007/JHEP12(2016)147} {\bibfield  {journal}
  {\bibinfo  {journal} {JHEP}\ }\textbf {\bibinfo {volume} {12}},\ \bibinfo
  {pages} {147}},\ \Eprint {https://arxiv.org/abs/1609.06203} {arXiv:1609.06203
  [hep-th]} \BibitemShut {NoStop}%
\bibitem [{\citenamefont {Donnay}\ \emph {et~al.}(2022)\citenamefont {Donnay},
  \citenamefont {Fiorucci}, \citenamefont {Herfray},\ and\ \citenamefont
  {Ruzziconi}}]{Donnay:2022aba}%
  \BibitemOpen
  \bibfield  {author} {\bibinfo {author} {\bibfnamefont {L.}~\bibnamefont
  {Donnay}}, \bibinfo {author} {\bibfnamefont {A.}~\bibnamefont {Fiorucci}},
  \bibinfo {author} {\bibfnamefont {Y.}~\bibnamefont {Herfray}},\ and\ \bibinfo
  {author} {\bibfnamefont {R.}~\bibnamefont {Ruzziconi}},\ }\bibfield  {title}
  {\bibinfo {title} {{Carrollian Perspective on Celestial Holography}},\ }\href
  {https://doi.org/10.1103/PhysRevLett.129.071602} {\bibfield  {journal}
  {\bibinfo  {journal} {Phys. Rev. Lett.}\ }\textbf {\bibinfo {volume} {129}},\
  \bibinfo {pages} {071602} (\bibinfo {year} {2022})},\ \Eprint
  {https://arxiv.org/abs/2202.04702} {arXiv:2202.04702 [hep-th]} \BibitemShut
  {NoStop}%
\bibitem [{\citenamefont {Bagchi}\ \emph
  {et~al.}(2022{\natexlab{a}})\citenamefont {Bagchi}, \citenamefont {Banerjee},
  \citenamefont {Basu},\ and\ \citenamefont {Dutta}}]{Bagchi:2022emh}%
  \BibitemOpen
  \bibfield  {author} {\bibinfo {author} {\bibfnamefont {A.}~\bibnamefont
  {Bagchi}}, \bibinfo {author} {\bibfnamefont {S.}~\bibnamefont {Banerjee}},
  \bibinfo {author} {\bibfnamefont {R.}~\bibnamefont {Basu}},\ and\ \bibinfo
  {author} {\bibfnamefont {S.}~\bibnamefont {Dutta}},\ }\bibfield  {title}
  {\bibinfo {title} {{Scattering Amplitudes: Celestial and Carrollian}},\
  }\href {https://doi.org/10.1103/PhysRevLett.128.241601} {\bibfield  {journal}
  {\bibinfo  {journal} {Phys. Rev. Lett.}\ }\textbf {\bibinfo {volume} {128}},\
  \bibinfo {pages} {241601} (\bibinfo {year} {2022}{\natexlab{a}})},\ \Eprint
  {https://arxiv.org/abs/2202.08438} {arXiv:2202.08438 [hep-th]} \BibitemShut
  {NoStop}%
\bibitem [{\citenamefont {Donnay}\ \emph {et~al.}(2023)\citenamefont {Donnay},
  \citenamefont {Fiorucci}, \citenamefont {Herfray},\ and\ \citenamefont
  {Ruzziconi}}]{Donnay:2022wvx}%
  \BibitemOpen
  \bibfield  {author} {\bibinfo {author} {\bibfnamefont {L.}~\bibnamefont
  {Donnay}}, \bibinfo {author} {\bibfnamefont {A.}~\bibnamefont {Fiorucci}},
  \bibinfo {author} {\bibfnamefont {Y.}~\bibnamefont {Herfray}},\ and\ \bibinfo
  {author} {\bibfnamefont {R.}~\bibnamefont {Ruzziconi}},\ }\bibfield  {title}
  {\bibinfo {title} {{Bridging Carrollian and celestial holography}},\ }\href
  {https://doi.org/10.1103/PhysRevD.107.126027} {\bibfield  {journal} {\bibinfo
   {journal} {Phys. Rev. D}\ }\textbf {\bibinfo {volume} {107}},\ \bibinfo
  {pages} {126027} (\bibinfo {year} {2023})},\ \Eprint
  {https://arxiv.org/abs/2212.12553} {arXiv:2212.12553 [hep-th]} \BibitemShut
  {NoStop}%
\bibitem [{\citenamefont {Bagchi}\ \emph
  {et~al.}(2023{\natexlab{b}})\citenamefont {Bagchi}, \citenamefont
  {Dhivakar},\ and\ \citenamefont {Dutta}}]{Bagchi:2023fbj}%
  \BibitemOpen
  \bibfield  {author} {\bibinfo {author} {\bibfnamefont {A.}~\bibnamefont
  {Bagchi}}, \bibinfo {author} {\bibfnamefont {P.}~\bibnamefont {Dhivakar}},\
  and\ \bibinfo {author} {\bibfnamefont {S.}~\bibnamefont {Dutta}},\ }\bibfield
   {title} {\bibinfo {title} {{AdS Witten diagrams to Carrollian
  correlators}},\ }\href {https://doi.org/10.1007/JHEP04(2023)135} {\bibfield
  {journal} {\bibinfo  {journal} {JHEP}\ }\textbf {\bibinfo {volume} {04}},\
  \bibinfo {pages} {135}},\ \Eprint {https://arxiv.org/abs/2303.07388}
  {arXiv:2303.07388 [hep-th]} \BibitemShut {NoStop}%
\bibitem [{\citenamefont {Saha}(2023)}]{Saha:2023hsl}%
  \BibitemOpen
  \bibfield  {author} {\bibinfo {author} {\bibfnamefont {A.}~\bibnamefont
  {Saha}},\ }\bibfield  {title} {\bibinfo {title} {{Carrollian approach to 1 +
  3D flat holography}},\ }\href {https://doi.org/10.1007/JHEP06(2023)051}
  {\bibfield  {journal} {\bibinfo  {journal} {JHEP}\ }\textbf {\bibinfo
  {volume} {06}},\ \bibinfo {pages} {051}},\ \Eprint
  {https://arxiv.org/abs/2304.02696} {arXiv:2304.02696 [hep-th]} \BibitemShut
  {NoStop}%
\bibitem [{\citenamefont {Nguyen}\ and\ \citenamefont
  {West}(2023)}]{Nguyen:2023vfz}%
  \BibitemOpen
  \bibfield  {author} {\bibinfo {author} {\bibfnamefont {K.}~\bibnamefont
  {Nguyen}}\ and\ \bibinfo {author} {\bibfnamefont {P.}~\bibnamefont {West}},\
  }\bibfield  {title} {\bibinfo {title} {{Carrollian conformal fields and flat
  holography}},\ }\href@noop {} {\  (\bibinfo {year} {2023})},\ \Eprint
  {https://arxiv.org/abs/2305.02884} {arXiv:2305.02884 [hep-th]} \BibitemShut
  {NoStop}%
\bibitem [{\citenamefont {Bagchi}\ \emph
  {et~al.}(2023{\natexlab{c}})\citenamefont {Bagchi}, \citenamefont {Banerjee},
  \citenamefont {Basu}, \citenamefont {Islam},\ and\ \citenamefont
  {Mondal}}]{Bagchi:2022eui}%
  \BibitemOpen
  \bibfield  {author} {\bibinfo {author} {\bibfnamefont {A.}~\bibnamefont
  {Bagchi}}, \bibinfo {author} {\bibfnamefont {A.}~\bibnamefont {Banerjee}},
  \bibinfo {author} {\bibfnamefont {R.}~\bibnamefont {Basu}}, \bibinfo {author}
  {\bibfnamefont {M.}~\bibnamefont {Islam}},\ and\ \bibinfo {author}
  {\bibfnamefont {S.}~\bibnamefont {Mondal}},\ }\bibfield  {title} {\bibinfo
  {title} {{Magic fermions: Carroll and flat bands}},\ }\href
  {https://doi.org/10.1007/JHEP03(2023)227} {\bibfield  {journal} {\bibinfo
  {journal} {JHEP}\ }\textbf {\bibinfo {volume} {03}},\ \bibinfo {pages}
  {227}},\ \Eprint {https://arxiv.org/abs/2211.11640} {arXiv:2211.11640
  [hep-th]} \BibitemShut {NoStop}%
\bibitem [{\citenamefont {Bidussi}\ \emph {et~al.}(2022)\citenamefont
  {Bidussi}, \citenamefont {Hartong}, \citenamefont {Have}, \citenamefont
  {Musaeus},\ and\ \citenamefont {Prohazka}}]{Bidussi:2021nmp}%
  \BibitemOpen
  \bibfield  {author} {\bibinfo {author} {\bibfnamefont {L.}~\bibnamefont
  {Bidussi}}, \bibinfo {author} {\bibfnamefont {J.}~\bibnamefont {Hartong}},
  \bibinfo {author} {\bibfnamefont {E.}~\bibnamefont {Have}}, \bibinfo {author}
  {\bibfnamefont {J.}~\bibnamefont {Musaeus}},\ and\ \bibinfo {author}
  {\bibfnamefont {S.}~\bibnamefont {Prohazka}},\ }\bibfield  {title} {\bibinfo
  {title} {{Fractons, dipole symmetries and curved spacetime}},\ }\href
  {https://doi.org/10.21468/SciPostPhys.12.6.205} {\bibfield  {journal}
  {\bibinfo  {journal} {SciPost Phys.}\ }\textbf {\bibinfo {volume} {12}},\
  \bibinfo {pages} {205} (\bibinfo {year} {2022})},\ \Eprint
  {https://arxiv.org/abs/2111.03668} {arXiv:2111.03668 [hep-th]} \BibitemShut
  {NoStop}%
\bibitem [{\citenamefont {de~Boer}\ \emph {et~al.}(2022)\citenamefont
  {de~Boer}, \citenamefont {Hartong}, \citenamefont {Obers}, \citenamefont
  {Sybesma},\ and\ \citenamefont {Vandoren}}]{deBoer:2021jej}%
  \BibitemOpen
  \bibfield  {author} {\bibinfo {author} {\bibfnamefont {J.}~\bibnamefont
  {de~Boer}}, \bibinfo {author} {\bibfnamefont {J.}~\bibnamefont {Hartong}},
  \bibinfo {author} {\bibfnamefont {N.~A.}\ \bibnamefont {Obers}}, \bibinfo
  {author} {\bibfnamefont {W.}~\bibnamefont {Sybesma}},\ and\ \bibinfo {author}
  {\bibfnamefont {S.}~\bibnamefont {Vandoren}},\ }\bibfield  {title} {\bibinfo
  {title} {{Carroll Symmetry, Dark Energy and Inflation}},\ }\href
  {https://doi.org/10.3389/fphy.2022.810405} {\bibfield  {journal} {\bibinfo
  {journal} {Front. in Phys.}\ }\textbf {\bibinfo {volume} {10}},\ \bibinfo
  {pages} {810405} (\bibinfo {year} {2022})},\ \Eprint
  {https://arxiv.org/abs/2110.02319} {arXiv:2110.02319 [hep-th]} \BibitemShut
  {NoStop}%
\bibitem [{\citenamefont {Bagchi}(2013)}]{Bagchi:2013bga}%
  \BibitemOpen
  \bibfield  {author} {\bibinfo {author} {\bibfnamefont {A.}~\bibnamefont
  {Bagchi}},\ }\bibfield  {title} {\bibinfo {title} {{Tensionless Strings and
  Galilean Conformal Algebra}},\ }\href
  {https://doi.org/10.1007/JHEP05(2013)141} {\bibfield  {journal} {\bibinfo
  {journal} {JHEP}\ }\textbf {\bibinfo {volume} {05}},\ \bibinfo {pages}
  {141}},\ \Eprint {https://arxiv.org/abs/1303.0291} {arXiv:1303.0291 [hep-th]}
  \BibitemShut {NoStop}%
\bibitem [{\citenamefont {Bagchi}\ \emph
  {et~al.}(2016{\natexlab{b}})\citenamefont {Bagchi}, \citenamefont
  {Chakrabortty},\ and\ \citenamefont {Parekh}}]{Bagchi:2015nca}%
  \BibitemOpen
  \bibfield  {author} {\bibinfo {author} {\bibfnamefont {A.}~\bibnamefont
  {Bagchi}}, \bibinfo {author} {\bibfnamefont {S.}~\bibnamefont
  {Chakrabortty}},\ and\ \bibinfo {author} {\bibfnamefont {P.}~\bibnamefont
  {Parekh}},\ }\bibfield  {title} {\bibinfo {title} {{Tensionless Strings from
  Worldsheet Symmetries}},\ }\href {https://doi.org/10.1007/JHEP01(2016)158}
  {\bibfield  {journal} {\bibinfo  {journal} {JHEP}\ }\textbf {\bibinfo
  {volume} {01}},\ \bibinfo {pages} {158}},\ \Eprint
  {https://arxiv.org/abs/1507.04361} {arXiv:1507.04361 [hep-th]} \BibitemShut
  {NoStop}%
\bibitem [{\citenamefont {Bagchi}\ \emph {et~al.}(2020)\citenamefont {Bagchi},
  \citenamefont {Banerjee}, \citenamefont {Chakrabortty}, \citenamefont
  {Dutta},\ and\ \citenamefont {Parekh}}]{Bagchi:2020fpr}%
  \BibitemOpen
  \bibfield  {author} {\bibinfo {author} {\bibfnamefont {A.}~\bibnamefont
  {Bagchi}}, \bibinfo {author} {\bibfnamefont {A.}~\bibnamefont {Banerjee}},
  \bibinfo {author} {\bibfnamefont {S.}~\bibnamefont {Chakrabortty}}, \bibinfo
  {author} {\bibfnamefont {S.}~\bibnamefont {Dutta}},\ and\ \bibinfo {author}
  {\bibfnamefont {P.}~\bibnamefont {Parekh}},\ }\bibfield  {title} {\bibinfo
  {title} {{A tale of three \textemdash{} tensionless strings and vacuum
  structure}},\ }\href {https://doi.org/10.1007/JHEP04(2020)061} {\bibfield
  {journal} {\bibinfo  {journal} {JHEP}\ }\textbf {\bibinfo {volume} {04}},\
  \bibinfo {pages} {061}},\ \Eprint {https://arxiv.org/abs/2001.00354}
  {arXiv:2001.00354 [hep-th]} \BibitemShut {NoStop}%
\bibitem [{\citenamefont {Penna}(2018)}]{Penna:2018gfx}%
  \BibitemOpen
  \bibfield  {author} {\bibinfo {author} {\bibfnamefont {R.~F.}\ \bibnamefont
  {Penna}},\ }\bibfield  {title} {\bibinfo {title} {Near-horizon carroll
  symmetry and black hole love numbers},\ }\href@noop {} {\  (\bibinfo {year}
  {2018})},\ \Eprint {https://arxiv.org/abs/1812.05643} {arXiv:1812.05643
  [hep-th]} \BibitemShut {NoStop}%
\bibitem [{\citenamefont {Donnay}\ and\ \citenamefont
  {Marteau}(2019)}]{Donnay:2019jiz}%
  \BibitemOpen
  \bibfield  {author} {\bibinfo {author} {\bibfnamefont {L.}~\bibnamefont
  {Donnay}}\ and\ \bibinfo {author} {\bibfnamefont {C.}~\bibnamefont
  {Marteau}},\ }\bibfield  {title} {\bibinfo {title} {Carrollian physics at the
  black hole horizon},\ }\href {https://doi.org/10.1088/1361-6382/ab2fd5}
  {\bibfield  {journal} {\bibinfo  {journal} {Class. Quant. Grav.}\ }\textbf
  {\bibinfo {volume} {36}},\ \bibinfo {pages} {165002} (\bibinfo {year}
  {2019})},\ \Eprint {https://arxiv.org/abs/1903.09654} {arXiv:1903.09654
  [hep-th]} \BibitemShut {NoStop}%
\bibitem [{\citenamefont {Freidel}\ and\ \citenamefont
  {Jai-akson}(2022)}]{Freidel:2022vjq}%
  \BibitemOpen
  \bibfield  {author} {\bibinfo {author} {\bibfnamefont {L.}~\bibnamefont
  {Freidel}}\ and\ \bibinfo {author} {\bibfnamefont {P.}~\bibnamefont
  {Jai-akson}},\ }\bibfield  {title} {\bibinfo {title} {Carrollian
  hydrodynamics and symplectic structure on stretched horizons},\ }\href@noop
  {} {\  (\bibinfo {year} {2022})},\ \Eprint {https://arxiv.org/abs/2211.06415}
  {arXiv:2211.06415 [gr-qc]} \BibitemShut {NoStop}%
\bibitem [{\citenamefont {Redondo-Yuste}\ and\ \citenamefont
  {Lehner}(2023)}]{Redondo-Yuste:2022czg}%
  \BibitemOpen
  \bibfield  {author} {\bibinfo {author} {\bibfnamefont {J.}~\bibnamefont
  {Redondo-Yuste}}\ and\ \bibinfo {author} {\bibfnamefont {L.}~\bibnamefont
  {Lehner}},\ }\bibfield  {title} {\bibinfo {title} {Non-linear black hole
  dynamics and carrollian fluids},\ }\href
  {https://doi.org/10.1007/JHEP02(2023)240} {\bibfield  {journal} {\bibinfo
  {journal} {JHEP}\ }\textbf {\bibinfo {volume} {02}},\ \bibinfo {pages}
  {240}},\ \Eprint {https://arxiv.org/abs/2212.06175} {arXiv:2212.06175
  [gr-qc]} \BibitemShut {NoStop}%
\bibitem [{\citenamefont {Bagchi}\ \emph
  {et~al.}(2022{\natexlab{b}})\citenamefont {Bagchi}, \citenamefont
  {Grumiller},\ and\ \citenamefont {Sheikh-Jabbari}}]{Bagchi:2022iqb}%
  \BibitemOpen
  \bibfield  {author} {\bibinfo {author} {\bibfnamefont {A.}~\bibnamefont
  {Bagchi}}, \bibinfo {author} {\bibfnamefont {D.}~\bibnamefont {Grumiller}},\
  and\ \bibinfo {author} {\bibfnamefont {M.~M.}\ \bibnamefont
  {Sheikh-Jabbari}},\ }\bibfield  {title} {\bibinfo {title} {{Horizon Strings
  as 3d Black Hole Microstates}},\ }\href@noop {} {\  (\bibinfo {year}
  {2022}{\natexlab{b}})},\ \Eprint {https://arxiv.org/abs/2210.10794}
  {arXiv:2210.10794 [hep-th]} \BibitemShut {NoStop}%
\bibitem [{\citenamefont {de~Boer}\ \emph {et~al.}(2023)\citenamefont
  {de~Boer}, \citenamefont {Hartong}, \citenamefont {Obers}, \citenamefont
  {Sybesma},\ and\ \citenamefont {Vandoren}}]{deBoer:2023fnj}%
  \BibitemOpen
  \bibfield  {author} {\bibinfo {author} {\bibfnamefont {J.}~\bibnamefont
  {de~Boer}}, \bibinfo {author} {\bibfnamefont {J.}~\bibnamefont {Hartong}},
  \bibinfo {author} {\bibfnamefont {N.~A.}\ \bibnamefont {Obers}}, \bibinfo
  {author} {\bibfnamefont {W.}~\bibnamefont {Sybesma}},\ and\ \bibinfo {author}
  {\bibfnamefont {S.}~\bibnamefont {Vandoren}},\ }\bibfield  {title} {\bibinfo
  {title} {{Carroll stories}},\ }\href@noop {} {\  (\bibinfo {year} {2023})},\
  \Eprint {https://arxiv.org/abs/2307.06827} {arXiv:2307.06827 [hep-th]}
  \BibitemShut {NoStop}%
\bibitem [{\citenamefont {Ecker}\ \emph {et~al.}(2023)\citenamefont {Ecker},
  \citenamefont {Grumiller}, \citenamefont {Hartong}, \citenamefont {P\'erez},
  \citenamefont {Prohazka},\ and\ \citenamefont {Troncoso}}]{Ecker:2023uwm}%
  \BibitemOpen
  \bibfield  {author} {\bibinfo {author} {\bibfnamefont {F.}~\bibnamefont
  {Ecker}}, \bibinfo {author} {\bibfnamefont {D.}~\bibnamefont {Grumiller}},
  \bibinfo {author} {\bibfnamefont {J.}~\bibnamefont {Hartong}}, \bibinfo
  {author} {\bibfnamefont {A.}~\bibnamefont {P\'erez}}, \bibinfo {author}
  {\bibfnamefont {S.}~\bibnamefont {Prohazka}},\ and\ \bibinfo {author}
  {\bibfnamefont {R.}~\bibnamefont {Troncoso}},\ }\bibfield  {title} {\bibinfo
  {title} {{Carroll black holes}},\ }\href@noop {} {\  (\bibinfo {year}
  {2023})},\ \Eprint {https://arxiv.org/abs/2308.10947} {arXiv:2308.10947
  [hep-th]} \BibitemShut {NoStop}%
\bibitem [{\citenamefont {Gubser}\ and\ \citenamefont
  {Yarom}(2011)}]{Gubser:2010ui}%
  \BibitemOpen
  \bibfield  {author} {\bibinfo {author} {\bibfnamefont {S.~S.}\ \bibnamefont
  {Gubser}}\ and\ \bibinfo {author} {\bibfnamefont {A.}~\bibnamefont {Yarom}},\
  }\bibfield  {title} {\bibinfo {title} {{Conformal hydrodynamics in Minkowski
  and de Sitter spacetimes}},\ }\href
  {https://doi.org/10.1016/j.nuclphysb.2011.01.012} {\bibfield  {journal}
  {\bibinfo  {journal} {Nucl. Phys. B}\ }\textbf {\bibinfo {volume} {846}},\
  \bibinfo {pages} {469} (\bibinfo {year} {2011})},\ \Eprint
  {https://arxiv.org/abs/1012.1314} {arXiv:1012.1314 [hep-th]} \BibitemShut
  {NoStop}%
\bibitem [{\citenamefont {Armas}\ and\ \citenamefont
  {Have}(2023)}]{Armas:2023dcz}%
  \BibitemOpen
  \bibfield  {author} {\bibinfo {author} {\bibfnamefont {J.}~\bibnamefont
  {Armas}}\ and\ \bibinfo {author} {\bibfnamefont {E.}~\bibnamefont {Have}},\
  }\bibfield  {title} {\bibinfo {title} {{Carrollian fluids and spontaneous
  breaking of boost symmetry}},\ }\href@noop {} {\  (\bibinfo {year} {2023})},\
  \Eprint {https://arxiv.org/abs/2308.10594} {arXiv:2308.10594 [hep-th]}
  \BibitemShut {NoStop}%
\bibitem [{\citenamefont {Gibbons}\ \emph {et~al.}(2009)\citenamefont
  {Gibbons}, \citenamefont {Herdeiro}, \citenamefont {Warnick},\ and\
  \citenamefont {Werner}}]{Gibbons:2008zi}%
  \BibitemOpen
  \bibfield  {author} {\bibinfo {author} {\bibfnamefont {G.~W.}\ \bibnamefont
  {Gibbons}}, \bibinfo {author} {\bibfnamefont {C.~A.~R.}\ \bibnamefont
  {Herdeiro}}, \bibinfo {author} {\bibfnamefont {C.~M.}\ \bibnamefont
  {Warnick}},\ and\ \bibinfo {author} {\bibfnamefont {M.~C.}\ \bibnamefont
  {Werner}},\ }\bibfield  {title} {\bibinfo {title} {{Stationary Metrics and
  Optical Zermelo-Randers-Finsler Geometry}},\ }\href
  {https://doi.org/10.1103/PhysRevD.79.044022} {\bibfield  {journal} {\bibinfo
  {journal} {Phys. Rev. D}\ }\textbf {\bibinfo {volume} {79}},\ \bibinfo
  {pages} {044022} (\bibinfo {year} {2009})},\ \Eprint
  {https://arxiv.org/abs/0811.2877} {arXiv:0811.2877 [gr-qc]} \BibitemShut
  {NoStop}%
\bibitem [{\citenamefont {Damour}(1978)}]{PhysRevD.18.3598}%
  \BibitemOpen
  \bibfield  {author} {\bibinfo {author} {\bibfnamefont {T.}~\bibnamefont
  {Damour}},\ }\bibfield  {title} {\bibinfo {title} {Black-hole eddy
  currents},\ }\href {https://doi.org/10.1103/PhysRevD.18.3598} {\bibfield
  {journal} {\bibinfo  {journal} {Phys. Rev. D}\ }\textbf {\bibinfo {volume}
  {18}},\ \bibinfo {pages} {3598} (\bibinfo {year} {1978})}\BibitemShut
  {NoStop}%
\bibitem [{\citenamefont {Price}\ and\ \citenamefont
  {Thorne}(1986)}]{PhysRevD.33.915}%
  \BibitemOpen
  \bibfield  {author} {\bibinfo {author} {\bibfnamefont {R.~H.}\ \bibnamefont
  {Price}}\ and\ \bibinfo {author} {\bibfnamefont {K.~S.}\ \bibnamefont
  {Thorne}},\ }\bibfield  {title} {\bibinfo {title} {Membrane viewpoint on
  black holes: Properties and evolution of the stretched horizon},\ }\href
  {https://doi.org/10.1103/PhysRevD.33.915} {\bibfield  {journal} {\bibinfo
  {journal} {Phys. Rev. D}\ }\textbf {\bibinfo {volume} {33}},\ \bibinfo
  {pages} {915} (\bibinfo {year} {1986})}\BibitemShut {NoStop}%
\bibitem [{\citenamefont {Bhattacharyya}\ \emph
  {et~al.}(2008{\natexlab{a}})\citenamefont {Bhattacharyya}, \citenamefont
  {Hubeny}, \citenamefont {Minwalla},\ and\ \citenamefont
  {Rangamani}}]{Bhattacharyya:2007vjd}%
  \BibitemOpen
  \bibfield  {author} {\bibinfo {author} {\bibfnamefont {S.}~\bibnamefont
  {Bhattacharyya}}, \bibinfo {author} {\bibfnamefont {V.~E.}\ \bibnamefont
  {Hubeny}}, \bibinfo {author} {\bibfnamefont {S.}~\bibnamefont {Minwalla}},\
  and\ \bibinfo {author} {\bibfnamefont {M.}~\bibnamefont {Rangamani}},\
  }\bibfield  {title} {\bibinfo {title} {{Nonlinear Fluid Dynamics from
  Gravity}},\ }\href {https://doi.org/10.1088/1126-6708/2008/02/045} {\bibfield
   {journal} {\bibinfo  {journal} {JHEP}\ }\textbf {\bibinfo {volume} {02}},\
  \bibinfo {pages} {045}},\ \Eprint {https://arxiv.org/abs/0712.2456}
  {arXiv:0712.2456 [hep-th]} \BibitemShut {NoStop}%
\bibitem [{\citenamefont {Bhattacharyya}\ \emph
  {et~al.}(2008{\natexlab{b}})\citenamefont {Bhattacharyya}, \citenamefont
  {Loganayagam}, \citenamefont {Mandal}, \citenamefont {Minwalla},\ and\
  \citenamefont {Sharma}}]{Bhattacharyya:2008mz}%
  \BibitemOpen
  \bibfield  {author} {\bibinfo {author} {\bibfnamefont {S.}~\bibnamefont
  {Bhattacharyya}}, \bibinfo {author} {\bibfnamefont {R.}~\bibnamefont
  {Loganayagam}}, \bibinfo {author} {\bibfnamefont {I.}~\bibnamefont {Mandal}},
  \bibinfo {author} {\bibfnamefont {S.}~\bibnamefont {Minwalla}},\ and\
  \bibinfo {author} {\bibfnamefont {A.}~\bibnamefont {Sharma}},\ }\bibfield
  {title} {\bibinfo {title} {{Conformal Nonlinear Fluid Dynamics from Gravity
  in Arbitrary Dimensions}},\ }\href
  {https://doi.org/10.1088/1126-6708/2008/12/116} {\bibfield  {journal}
  {\bibinfo  {journal} {JHEP}\ }\textbf {\bibinfo {volume} {12}},\ \bibinfo
  {pages} {116}},\ \Eprint {https://arxiv.org/abs/0809.4272} {arXiv:0809.4272
  [hep-th]} \BibitemShut {NoStop}%
\bibitem [{\citenamefont {Janik}\ and\ \citenamefont
  {Peschanski}(2006{\natexlab{a}})}]{Janik:2005zt}%
  \BibitemOpen
  \bibfield  {author} {\bibinfo {author} {\bibfnamefont {R.~A.}\ \bibnamefont
  {Janik}}\ and\ \bibinfo {author} {\bibfnamefont {R.~B.}\ \bibnamefont
  {Peschanski}},\ }\bibfield  {title} {\bibinfo {title} {{Asymptotic perfect
  fluid dynamics as a consequence of Ads/CFT}},\ }\href
  {https://doi.org/10.1103/PhysRevD.73.045013} {\bibfield  {journal} {\bibinfo
  {journal} {Phys. Rev. D}\ }\textbf {\bibinfo {volume} {73}},\ \bibinfo
  {pages} {045013} (\bibinfo {year} {2006}{\natexlab{a}})},\ \Eprint
  {https://arxiv.org/abs/hep-th/0512162} {arXiv:hep-th/0512162} \BibitemShut
  {NoStop}%
\bibitem [{\citenamefont {Janik}\ and\ \citenamefont
  {Peschanski}(2006{\natexlab{b}})}]{Janik:2006gp}%
  \BibitemOpen
  \bibfield  {author} {\bibinfo {author} {\bibfnamefont {R.~A.}\ \bibnamefont
  {Janik}}\ and\ \bibinfo {author} {\bibfnamefont {R.~B.}\ \bibnamefont
  {Peschanski}},\ }\bibfield  {title} {\bibinfo {title} {{Gauge/gravity duality
  and thermalization of a boost-invariant perfect fluid}},\ }\href
  {https://doi.org/10.1103/PhysRevD.74.046007} {\bibfield  {journal} {\bibinfo
  {journal} {Phys. Rev. D}\ }\textbf {\bibinfo {volume} {74}},\ \bibinfo
  {pages} {046007} (\bibinfo {year} {2006}{\natexlab{b}})},\ \Eprint
  {https://arxiv.org/abs/hep-th/0606149} {arXiv:hep-th/0606149} \BibitemShut
  {NoStop}%
\bibitem [{\citenamefont {Janik}(2007)}]{Janik:2006ft}%
  \BibitemOpen
  \bibfield  {author} {\bibinfo {author} {\bibfnamefont {R.~A.}\ \bibnamefont
  {Janik}},\ }\bibfield  {title} {\bibinfo {title} {{Viscous plasma evolution
  from gravity using AdS/CFT}},\ }\href
  {https://doi.org/10.1103/PhysRevLett.98.022302} {\bibfield  {journal}
  {\bibinfo  {journal} {Phys. Rev. Lett.}\ }\textbf {\bibinfo {volume} {98}},\
  \bibinfo {pages} {022302} (\bibinfo {year} {2007})},\ \Eprint
  {https://arxiv.org/abs/hep-th/0610144} {arXiv:hep-th/0610144} \BibitemShut
  {NoStop}%
\bibitem [{\citenamefont {Grumiller}\ and\ \citenamefont
  {Romatschke}(2008)}]{Grumiller:2008va}%
  \BibitemOpen
  \bibfield  {author} {\bibinfo {author} {\bibfnamefont {D.}~\bibnamefont
  {Grumiller}}\ and\ \bibinfo {author} {\bibfnamefont {P.}~\bibnamefont
  {Romatschke}},\ }\bibfield  {title} {\bibinfo {title} {{On the collision of
  two shock waves in AdS(5)}},\ }\href
  {https://doi.org/10.1088/1126-6708/2008/08/027} {\bibfield  {journal}
  {\bibinfo  {journal} {JHEP}\ }\textbf {\bibinfo {volume} {08}},\ \bibinfo
  {pages} {027}},\ \Eprint {https://arxiv.org/abs/0803.3226} {arXiv:0803.3226
  [hep-th]} \BibitemShut {NoStop}%
\bibitem [{\citenamefont {Albacete}\ \emph {et~al.}(2008)\citenamefont
  {Albacete}, \citenamefont {Kovchegov},\ and\ \citenamefont
  {Taliotis}}]{Albacete:2008vs}%
  \BibitemOpen
  \bibfield  {author} {\bibinfo {author} {\bibfnamefont {J.~L.}\ \bibnamefont
  {Albacete}}, \bibinfo {author} {\bibfnamefont {Y.~V.}\ \bibnamefont
  {Kovchegov}},\ and\ \bibinfo {author} {\bibfnamefont {A.}~\bibnamefont
  {Taliotis}},\ }\bibfield  {title} {\bibinfo {title} {{Modeling Heavy Ion
  Collisions in AdS/CFT}},\ }\href
  {https://doi.org/10.1088/1126-6708/2008/07/100} {\bibfield  {journal}
  {\bibinfo  {journal} {JHEP}\ }\textbf {\bibinfo {volume} {07}},\ \bibinfo
  {pages} {100}},\ \Eprint {https://arxiv.org/abs/0805.2927} {arXiv:0805.2927
  [hep-th]} \BibitemShut {NoStop}%
\bibitem [{\citenamefont {Heller}\ \emph {et~al.}(2009)\citenamefont {Heller},
  \citenamefont {Surowka}, \citenamefont {Loganayagam}, \citenamefont
  {Spalinski},\ and\ \citenamefont {Vazquez}}]{Heller:2008mb}%
  \BibitemOpen
  \bibfield  {author} {\bibinfo {author} {\bibfnamefont {M.~P.}\ \bibnamefont
  {Heller}}, \bibinfo {author} {\bibfnamefont {P.}~\bibnamefont {Surowka}},
  \bibinfo {author} {\bibfnamefont {R.}~\bibnamefont {Loganayagam}}, \bibinfo
  {author} {\bibfnamefont {M.}~\bibnamefont {Spalinski}},\ and\ \bibinfo
  {author} {\bibfnamefont {S.~E.}\ \bibnamefont {Vazquez}},\ }\bibfield
  {title} {\bibinfo {title} {{Consistent Holographic Description of
  Boost-Invariant Plasma}},\ }\href
  {https://doi.org/10.1103/PhysRevLett.102.041601} {\bibfield  {journal}
  {\bibinfo  {journal} {Phys. Rev. Lett.}\ }\textbf {\bibinfo {volume} {102}},\
  \bibinfo {pages} {041601} (\bibinfo {year} {2009})},\ \Eprint
  {https://arxiv.org/abs/0805.3774} {arXiv:0805.3774 [hep-th]} \BibitemShut
  {NoStop}%
\bibitem [{\citenamefont {Kinoshita}\ \emph {et~al.}(2009)\citenamefont
  {Kinoshita}, \citenamefont {Mukohyama}, \citenamefont {Nakamura},\ and\
  \citenamefont {Oda}}]{Kinoshita:2008dq}%
  \BibitemOpen
  \bibfield  {author} {\bibinfo {author} {\bibfnamefont {S.}~\bibnamefont
  {Kinoshita}}, \bibinfo {author} {\bibfnamefont {S.}~\bibnamefont
  {Mukohyama}}, \bibinfo {author} {\bibfnamefont {S.}~\bibnamefont
  {Nakamura}},\ and\ \bibinfo {author} {\bibfnamefont {K.-y.}\ \bibnamefont
  {Oda}},\ }\bibfield  {title} {\bibinfo {title} {{A Holographic Dual of
  Bjorken Flow}},\ }\href {https://doi.org/10.1143/PTP.121.121} {\bibfield
  {journal} {\bibinfo  {journal} {Prog. Theor. Phys.}\ }\textbf {\bibinfo
  {volume} {121}},\ \bibinfo {pages} {121} (\bibinfo {year} {2009})},\ \Eprint
  {https://arxiv.org/abs/0807.3797} {arXiv:0807.3797 [hep-th]} \BibitemShut
  {NoStop}%
\bibitem [{\citenamefont {Beuf}\ \emph {et~al.}(2009)\citenamefont {Beuf},
  \citenamefont {Heller}, \citenamefont {Janik},\ and\ \citenamefont
  {Peschanski}}]{Beuf:2009cx}%
  \BibitemOpen
  \bibfield  {author} {\bibinfo {author} {\bibfnamefont {G.}~\bibnamefont
  {Beuf}}, \bibinfo {author} {\bibfnamefont {M.~P.}\ \bibnamefont {Heller}},
  \bibinfo {author} {\bibfnamefont {R.~A.}\ \bibnamefont {Janik}},\ and\
  \bibinfo {author} {\bibfnamefont {R.}~\bibnamefont {Peschanski}},\ }\bibfield
   {title} {\bibinfo {title} {{Boost-invariant early time dynamics from
  AdS/CFT}},\ }\href {https://doi.org/10.1088/1126-6708/2009/10/043} {\bibfield
   {journal} {\bibinfo  {journal} {JHEP}\ }\textbf {\bibinfo {volume} {10}},\
  \bibinfo {pages} {043}},\ \Eprint {https://arxiv.org/abs/0906.4423}
  {arXiv:0906.4423 [hep-th]} \BibitemShut {NoStop}%
\bibitem [{\citenamefont {Chesler}\ and\ \citenamefont
  {Yaffe}(2010)}]{Chesler:2009cy}%
  \BibitemOpen
  \bibfield  {author} {\bibinfo {author} {\bibfnamefont {P.~M.}\ \bibnamefont
  {Chesler}}\ and\ \bibinfo {author} {\bibfnamefont {L.~G.}\ \bibnamefont
  {Yaffe}},\ }\bibfield  {title} {\bibinfo {title} {{Boost invariant flow,
  black hole formation, and far-from-equilibrium dynamics in N = 4
  supersymmetric Yang-Mills theory}},\ }\href
  {https://doi.org/10.1103/PhysRevD.82.026006} {\bibfield  {journal} {\bibinfo
  {journal} {Phys. Rev. D}\ }\textbf {\bibinfo {volume} {82}},\ \bibinfo
  {pages} {026006} (\bibinfo {year} {2010})},\ \Eprint
  {https://arxiv.org/abs/0906.4426} {arXiv:0906.4426 [hep-th]} \BibitemShut
  {NoStop}%
\bibitem [{\citenamefont {Taliotis}(2010)}]{Taliotis:2010pi}%
  \BibitemOpen
  \bibfield  {author} {\bibinfo {author} {\bibfnamefont {A.}~\bibnamefont
  {Taliotis}},\ }\bibfield  {title} {\bibinfo {title} {{Heavy Ion Collisions
  with Transverse Dynamics from Evolving AdS Geometries}},\ }\href
  {https://doi.org/10.1007/JHEP09(2010)102} {\bibfield  {journal} {\bibinfo
  {journal} {JHEP}\ }\textbf {\bibinfo {volume} {09}},\ \bibinfo {pages}
  {102}},\ \Eprint {https://arxiv.org/abs/1004.3500} {arXiv:1004.3500 [hep-th]}
  \BibitemShut {NoStop}%
\end{thebibliography}%

\end{document}